\documentclass[onecolumn,12pt]{article}
\usepackage{jheppub}
\usepackage{ifpdf}
\usepackage[bb=boondox]{mathalfa}
\graphicspath{ {figures/} }

\usepackage{amsfonts}
\usepackage{latexsym}
\usepackage{color}

\usepackage{amsmath,braket}
\usepackage{amssymb,physics}
\usepackage{amsthm}
\usepackage{mathtools}

\usepackage{epsfig}

\usepackage{tensor}
\usepackage{pdfpages}
\usepackage{bbm}
\usepackage{graphicx,epstopdf}
\usepackage[numbers]{natbib} 
\usepackage[makeroom]{cancel}
\usepackage{hyperref}
\usepackage{array}
\usepackage[export]{adjustbox}

\usepackage[normalem]{ulem}
\usepackage{romannum}

\usepackage{ulem}

\usepackage{tikz}
\usetikzlibrary{intersections}
\usepackage{circuitikz}

\numberwithin{equation}{section}									

\usepackage{here}

\usepackage{comment}

\newcommand{\de}{\partial}
\newcommand{\be}{\begin{equation}}
\newcommand{\ba}{\begin{eqnarray}}
\newcommand{\ea}{\end{eqnarray}}
\newcommand{\ee}{\end{equation}}

\newcommand{\lr}{\leftrightarrow}

\newcommand{\s}{\sqrt}
\newcommand{\vp}{\varphi}

\newcommand{\ap}{\alpha}

\newcommand{\ddd}{\cdot\cdot\cdot}
\newcommand{\no}{\nonumber \\}
\newcommand{\la}{\langle}
\newcommand{\lb}{\rangle}
\newcommand{\bea}{\begin{eqnarray}}
	\newcommand{\eea}{\end{eqnarray}}
\newcommand{\bes}{\begin{equation*}}
	\newcommand{\beas}{\begin{eqnarray*}}
		\newcommand{\eeas}{\end{eqnarray*}}
	\newcommand{\bas}{\begin{array*}}
		\newcommand{\eas}{\end{array*}}
	\newcommand{\ees}{\end{equation*}}

\newcommand{\p}{\partial}
\newcommand{\ep}{\epsilon}

\let\a=\alpha    \let\e=\epsilon       
 \let\p=\phi \let\r=\rho 
\let\t=\tau   \let\vp=\varphi

\def\inf{\infty}

\newcommand{\mt}[1]{\textrm{\tiny #1}}

\newcommand{\con}{\mathrm{con}}
\newcommand{\dis}{\mathrm{dis}}

\renewcommand{\(}{\left(}
\renewcommand{\)}{\right)}

\newcommand{\GN}{G_\mt{N}}

\subheader{\today}
\title{\boldmath Flat Space Holography via AdS/BCFT}

\author[a]{Peng-Xiang Hao}
\author[a]{Naoki Ogawa}
\author[a,b]{Tadashi Takayanagi}
\author[a]{Takahiro Waki}

\affiliation[a]{Center\! for Gravitational\! Physics\! and Quantum \! Information, Yukawa\! Institute\! for\! Theoretical\! Physics, Kyoto\! University, Kitashirakawa\! Oiwakecho, Sakyo-ku, Kyoto 606-8502, Japan}
\affiliation[b]{Inamori\! Research\! Institute\! for\! Science,\! 620\! Suiginya-cho,\! Shimogyo-ku,\! Kyoto\! 600-8411, Japan}

\emailAdd{pxhao@yukawa.kyoto-u.ac.jp}
\emailAdd{naoki.ogawa@yukawa.kyoto-u.ac.jp}
\emailAdd{takayana@yukawa.kyoto-u.ac.jp}
\emailAdd{takahiro.waki@yukawa.kyoto-u.ac.jp}

\abstract{In this paper, we study a new class of AdS/BCFT setups, where the world-volumes of end-of-the-world branes (EOW branes) are given by flat spaces, to explore flat space holography from an AdS bulk. We show that they provide gravity duals of CFTs in the presence of null boundaries. Our holographic calculations lead to many new predictions on entanglement entropy, correlation functions and partition functions for CFTs with null boundaries. By considering a bulk region between two EOW branes, we present an AdS/BCFT explanation that the flat space gravity is dual to a Carrollian CFT (CCFT), including the swing surface calculation of entanglement entropy.}

\begin{document} 
	
\begin{flushright}
YITP-25-132
\\
\end{flushright}
\maketitle
\flushbottom

\section{Introduction}
\label{sec:intro}

The idea of holography \cite{tHooft:1993dmi,Susskind:1994vu} enables us to describe difficult problems in quantum gravity in terms of more tractable ones in dual quantum field theories. In particular, holography in anti de-Sitter space (AdS), i.e. the AdS/CFT \cite{Maldacena:1997re,Gubser:1998bc,Witten:1998qj}, has been well-established. In the AdS/CFT, the gravity on $d+1$ dimensional AdS becomes equivalent to a $d$ dimensional conformal field theory (CFT), which lives on the boundary of the AdS. This duality was found by considering a setup where D-branes in string theory back-react to the geometry via the gravitational force, which leads to the AdS geometry. Moreover, the $d$ dimensional conformal symmetry corresponds to the geometrical symmetry of $d+1$ dimensional AdS.

On the contrary, holography in flat spacetime turns out to be more difficult to obtain in string theory, because flat spacetime does not need any source in string theory, as opposed to the AdS which was produced by D-branes. 
However, since the holography typically argues that a gravity on a certain spacetime is dual to a field theory on its boundary, we expect that a gravity in a Minkowski spacetime should be dual to its null boundary. In addition, the asymptotic symmetries that preserve the boundary structure, so-called BMS group, are useful to explore the holography \cite{Bondi:1962px,Sachs:1962wk}.

So far, there are two major approaches to holography in asymptotically flat spacetimes. The first one is the Flat/Carrollian CFT (CCFT) correspondence \cite{Barnich:2010eb,Bagchi:2010zz,Fareghbal:2013ifa}, where the dual field theory resides on the null boundary. The second one is the celestial holography \cite{deBoer:2003vf,Pasterski:2016qvg,Pasterski:2017kqt}, where dual CFTs live on the celestial sphere which is a codimension two. The relationship between these two ideas of holography has been understood better till now: \cite{Donnay:2022wvx,Donnay:2022aba,Bagchi:2022emh,Bagchi:2023fbj}. In addition to these, which fully employ the symmetries, it would be desirable if we could have another argument which is a more top-down approach. 

The purpose of this paper is to provide a new approach to the flat space holography. In particular, we would like to consider a method to embed it in the framework of AdS/CFT. This is because it allows us to understand the flat space holography as a version of AdS/CFT, which is quite established and well understood. For this purpose, we employ the AdS/BCFT correspondence \cite{Takayanagi:2011zk,Fujita:2011fp,Karch:2000gx}, which argues that the gravity dual of a CFT on a $d$ dimensional manifold $N_d$ with a conformal boundary (BCFT) is dual to a part of AdS geometry $M_{d+1}$ which is surrounded by the manifold $N_d$ and a surface $Q_d$ called the end-of-the-world brane (EOW brane) such that $\de M_{d+1}=N_d\cup Q_d$. 

One advantage of the AdS/BCFT is that we can extend the holographic calculation of entanglement entropy \cite{Ryu:2006bv,Ryu:2006ef,Hubeny:2007xt} to the setups with EOW branes by simply allowing the extremal surface can end on EOW branes \cite{Takayanagi:2011zk,Fujita:2011fp}. Another advantage is that we can identify the setups of EOW branes in AdS geometries with the brane world models, namely Randall-Sundrum models, so called RS1 and RS2 model, \cite{Randall:1999ee,Randall:1999vf}. It has been known for a while that the brane world models enjoy an intriguing holographic duality, which generalizes the AdS/CFT
\cite{Gubser:1999vj,Giddings:2000mu,Arkani-Hamed:2000ijo,Rattazzi:2000hs,Perez-Victoria:2001lex,Soda:2010si}. This brane world holography argues that the classical gravity which lives the $d+1$ dimensional AdS bulk surrounded by the brane(s) is dual to a system where a $d$ dimensional CFT with a finite UV cut off is coupled to the $d$ dimensional dynamical quantum gravity. The gravity on the brane is dynamical because we impose the Neumann boundary on it. Notice that the $d$ dimensional gravity on the brane is known to be described by a modification of Einstein gravity \cite{Garriga:1999yh,Shiromizu:1999wj,Gubser:1999vj,Giddings:2000mu} and this modification is dual to the loop corrections to the graviton propagator due to the exchange of the CFT matter via the holography.  This brane world holography adds the third dual interpretation to the AdS/BCFT duality, which we will fully employ in this paper.

The EOW branes, the most important ingredient of the AdS/BCFT, are characterized by values of tension $T$. Depending on values of $T$, they are classified into three classes such that the induced metric on a given EOW brane becomes either AdS, dS or flat space. The standard one is the AdS brane, where gravity localizes on the brane \cite{Karch:2000ct}, called Karch-Randall model, and thus describe quantum gravity on AdS. This has successfully been applied to the black hole information problem \cite{Almheiri:2019hni,Almheiri:2019psf} as the holographic entanglement in the AdS/BCFT can explain the island formula \cite{Penington:2019npb,Almheiri:2019psf,Chen:2020uac}, which was explicitly confirmed in \cite{Suzuki:2022xwv,Izumi:2022opi}. 
For applications to cosmological models which employ the AdS/BCFT holography, refer to \cite{Cooper:2018cmb,Antonini:2019qkt,VanRaamsdonk:2021qgv,Omiya:2021olc,Antonini:2022blk,Waddell:2022fbn}. The holographic interpretations of dS branes have not been completely well-understood because they are exotic from the viewpoint of CFTs as they are dual to space-like boundaries in CFTs \cite{Akal:2020wfl,Chen:2020tes,Hao:2024nhd}. Nevertheless, several interesting applications  have been considered recently, including the gravity dual of cross cap states \cite{Wei:2024zez} and an embedding dS/CFT into AdS/BCFT \cite{Fujiki:2025yyf}.

From the AdS/BCFT viewpoint, the least studied EOW branes are the flat branes which are realized for special values of tension. However, we know that the flat space EOW brane leads to a holography for flat spacetimes, where the $d$ dimensional quantum gravity coupled with a CFT on a flat space gets equivalent to the $d+1$ dimensional bulk AdS surrounded by the brane(s), owing to the much earlier and many works on the brane-world holography, including \cite{Gubser:1999vj,Giddings:2000mu,Arkani-Hamed:2000ijo,Rattazzi:2000hs,Perez-Victoria:2001lex,Soda:2010si}. Motivated by this, the main purpose of this paper is explore the implications of holography for the flat EOW branes by combining the ideas of the modern AdS/BCFT with the celebrated brane-world holography in order to get insights on how the holography for flat spacetimes looks like. The main difference between our AdS/BCFT approach and the original brane-world holography is that in the former, the dual CFT always live on the UV boundary and the EOW brane is treated as the boundary condition in the IR, while in the latter, the finite cut off CFT lives on the brane. Our AdS/BCFT approach has the advantage that we can compute various quantities such as entanglement entropy and correlation functions in the fine grained CFT situated at the UV boundary of AdS, as we will see in this paper. Another intriguing point is that flat EOW branes are dual to null boundaries in CFTs. Such boundaries in field theories have not been well-understood till now and are worth exploring by combining the field theoretic and holographic calculations. 

Using a flat EOW brane, we will consider three setups of AdS/BCFT, called type I, II and III (refer to Fig.\ref{fig:setups}). In the type I and II model, the brane has a negative and positive tension, respectively. In the type I case, the CFT is defined on a diamond with null boundaries, where the boundaries play a role similar to the final state projections. On the other hand, in the type II case, the CFT on the diamond is coupled to flat space gravity through the null boundaries. In the type II construction, we will find that our AdS/BCFT consideration reproduces the  Flat/Carrollian CFT (CCFT) correspondence, where the Carrolian CFT is realized on the null boundary of the flat EOW brane. We can pick up purely the flat space gravity by sandwiching the bulk AdS region by two flat EOW branes and by applying the wedge holography \cite{Akal:2020wfl}, which is called the type III setup and can make the above argument clearer.  We would like to note that in the Poincare AdS coordinate patch, these three setups are not new and have been discussed by many authors. The type I model corresponds to the hard wall model with an IR cut off, which has often been used in AdS/QCD \cite{Erlich:2005qh}. The type II model and type III model are equivalent to the RS2 model \cite{Randall:1999vf} and RS1 model \cite{Randall:1999ee}, respectively. However, we would like to emphasize that in this paper we are mainly interested in the holography in the global AdS coordinate patch where we can always find the asymptotically AdS UV boundary, where we expect the dual fine grained CFT lives.

We will work out the holographic calculations of entanglement entropy and correlation functions, which are consistent with each other in the type I and II setup. In the type II case, we will see that the holographic entanglement includes an imaginary valued contribution, which can properly be regarded as
the pseudo entropy \cite{Nakata:2021ubr} and is similar to the time-like entanglement entropy \cite{Doi:2022iyj,Doi:2023zaf,Heller:2024whi}. Thus this implies that the dual density matrix is not hermitian. In the type III setup, we will derive the swing surface prescription of holographic entanglement in flat space holography \cite{Apolo:2020bld} from the AdS/BCFT. We will also discuss a Euclidean space version of flat space holography using the EOW brane, which implies that the gravity dual of a flat space is given by a point-like theory. 

This paper is organized as follows. In section two, we present our three setups of AdS/BCFT with flat EOW branes. In section three, we present the calculation of holographic entanglement entropy in our AdS/BCFT with flat EOW branes. In section four, we compute one-point and two-point functions in our setups. In section five, we evaluate the on-shell action of our gravity duals. In section six, we analyze the Euclidean counterpart of our AdS/BCFT with flat EOW branes. In section seven, we summarize our conclusions and discuss future problems. In appendix A, we present a field theoretic analysis of the limit of light-like boundary in a two dimensional BCFT.

{\it Note added}: When we are finalizing this work, we noticed the independent preprint \cite{Neuenfeld:2025wnl}, which discusses flat EOW branes from different viewpoints.

\section{AdS/BCFT Setups for Flat Space Holography}
\label{sec:setups}

A $d$ dimensional conformal field theory (CFT) defined on a manifold with boundaries with boundary conditions which preserve the boundary conformal symmetry $SO(2,d-1)$ of the whole symmetry $SO(2,d)$ is called the boundary conformal field theory (BCFT). The gravity dual of $d$ dimensional BCFT (BCFT$_d$)
is given by a part of AdS$_{d+1}$ surrounded by the end-of-the-world brane (EOW brane), described by the surface $Q$.  On this EOW brane, the following boundary condition is imposed so that the dual theory preserves the boundary conformal symmetry \cite{Takayanagi:2011zk,Fujita:2011fp}:
\ba \label{eq: brane eom}
K_{ab}-Kh_{ab}=-Th_{ab},
\ea
where $K_{ab}$ and $h_{ab}$ are extrinsic curvature and the induced metric of the EOW brane. The parameter 
$T$ is called the tension of the EOW brane, which determines the profile of the brane. We have the following three classes of EOW branes depending on the values of the tension:
\begin{align}
 \mbox{(i)} & \ \ |T|< \frac{d-1}{R}: \ \ \mbox{AdS brane in AdS$_{d+1}\lr$ Time-like boundary in CFT$_d$}\nonumber\\
 \mbox{(ii)} & \ \ |T|=\frac{d-1}{R}: \ \ \mbox{Flat brane in AdS$_{d+1}\lr $ Null boundary in CFT$_d$},\nonumber\\
 \mbox{(iii)} & \ \ |T|>\frac{d-1}{R}: \ \ \mbox{dS brane in AdS$_{d+1}\lr $ Space-like boundary in CFT$_d$}.\nonumber
\end{align}

We can further generalize each EOW brane by introducing the matter fields on the brane as was done in the AdS brane \cite{Kanda:2023jyi,Kanda:2023zse} and the dS brane \cite{Fujiki:2025yyf}, which breaks the boundary conformal invariance and thus can describe the boundary renormalization group flows. In particular, it would be intriguing to do this for our flat branes, though below we will be focusing on the case without any matter fields.

\subsection{Flat EOW brane}

The main purpose of this paper is to examine the second case (ii), which has not been well-studied in the context of holography till now. If we employ the brane-world holography idea to the AdS/BCFT, we can obtain the third interpretation in terms of the CFT on the $d$ dimensional manifold coupled to a $d$ dimensional gravity on the EOW brane along the boundaries, which is called the double holography and has been often employed to analyze the black hole information problems. If we apply this interpretation, we can regard the setup (ii) as a CFT coupled to a gravity on a  flat spacetime.  Therefore, we can embed the problem of the holography in flat spacetime, which is far from well-understood even now, into the AdS/BCFT setups, which is much more controllable. Refer also to the recent paper \cite{Fontanella:2025tbs} for another different approach, where the flat space limit is taken for the whole bulk AdS space.

\begin{figure}[ttt]
		\centering
		\includegraphics[width=10cm]{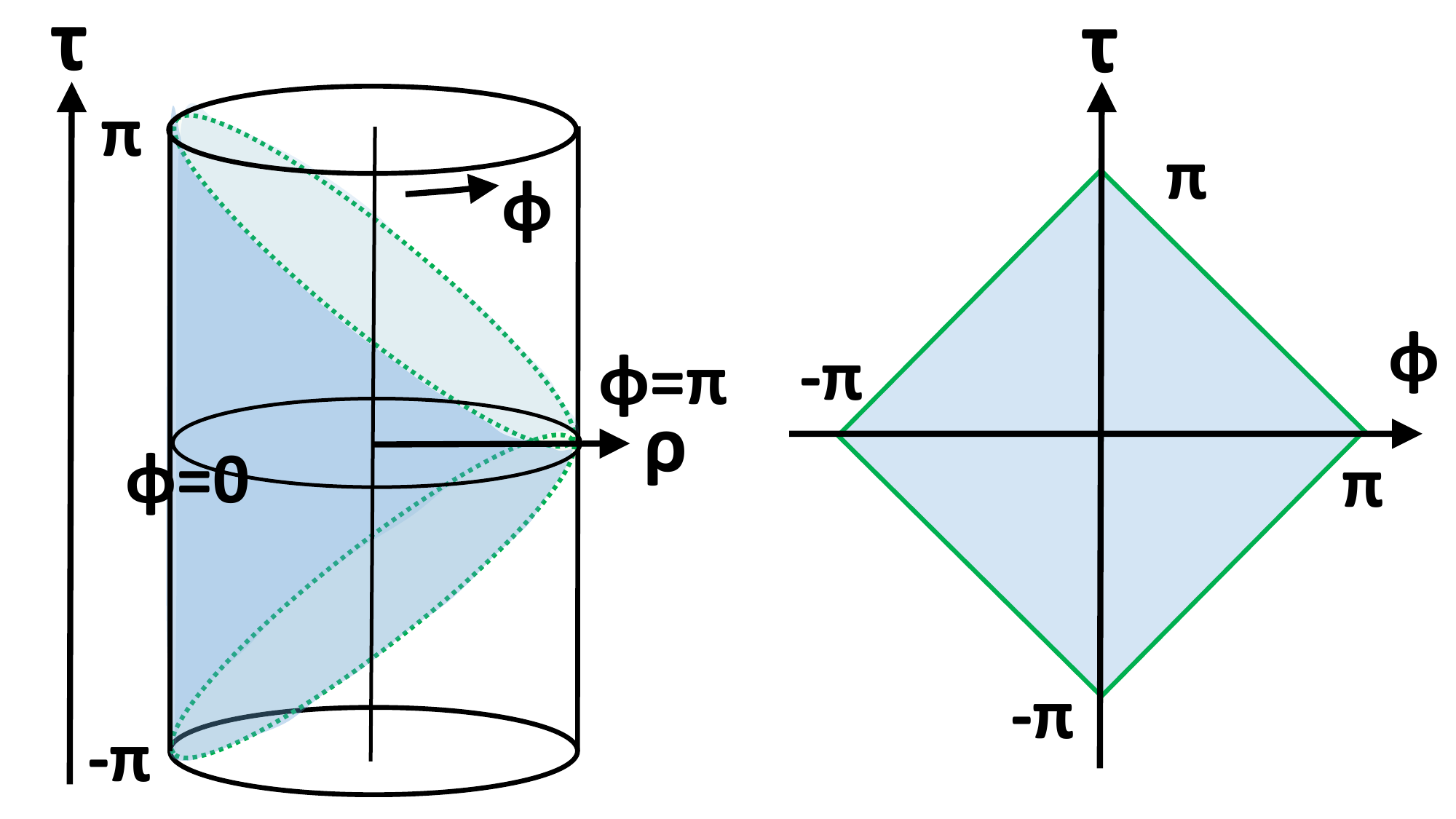}
		\caption{A sketch of Poincare AdS inside global AdS (left) and the boundary of Poincare AdS (right) in the coordinate of $(\tau,\phi)$.} 
		\label{fig:AdS}
\end{figure}

Now let us present how the flat brane (ii) looks like explicitly.  We consider both the Poincare and global AdS$_{d+1}$, whose metric looks like
\begin{align}
    ds^2&=R^2\left(\frac{dz^2-dt^2+\sum_{i=1}^{d-1}dx_i^2}{z^2}\right)\label{poinc}\nonumber\\
   &=R^2\left(-\cosh^2{\rho}d\tau^2+d\rho^2+\sinh^2{\rho}(d\phi^2+\sin^2\phi d\Omega^2)\right),
\end{align}
respectively. $(\Omega_1,\Omega_2,\ddd,\Omega_{d-1})$ describes the coordinate of a unit sphere $S^{d-2}$.
These two coordinates are related explicitly via
\begin{align}
\begin{aligned}\label{eq: embedding}
    X_0&=R\cosh{\rho}\cos{\tau}=\frac{z}{2}\left( 1+\frac{R^2+x^2-t^2}{z^2} \right),\\
    X_1&=R\cosh{\rho}\sin{\tau}=\frac{Rt}{z},\\
    X_2&=-R\sinh{\rho}\cos{\phi}=\frac{z}{2}\left( 1-\frac{R^2-x^2+t^2}{z^2} \right),\\
  X_{i+2}&=R\sinh{\rho}\sin{\phi}~\Omega_i=\frac{Rx_i}{z}, \ \ \ (i=1,2,\ddd,d-1).
\end{aligned}
\end{align}
Note that the Poincare coordinate only covers a wedge of global AdS because
\begin{align}
    \frac{X_0-X_2}{R}=\frac{R}{z}=\cosh{\rho}\cos{\tau}+\sinh{\rho}\cos{\phi}>0,
\end{align}
as depicted in the left panel of Fig.\ref{fig:AdS}. 

\begin{figure}[ttt]
		\centering
		\includegraphics[width=10cm]{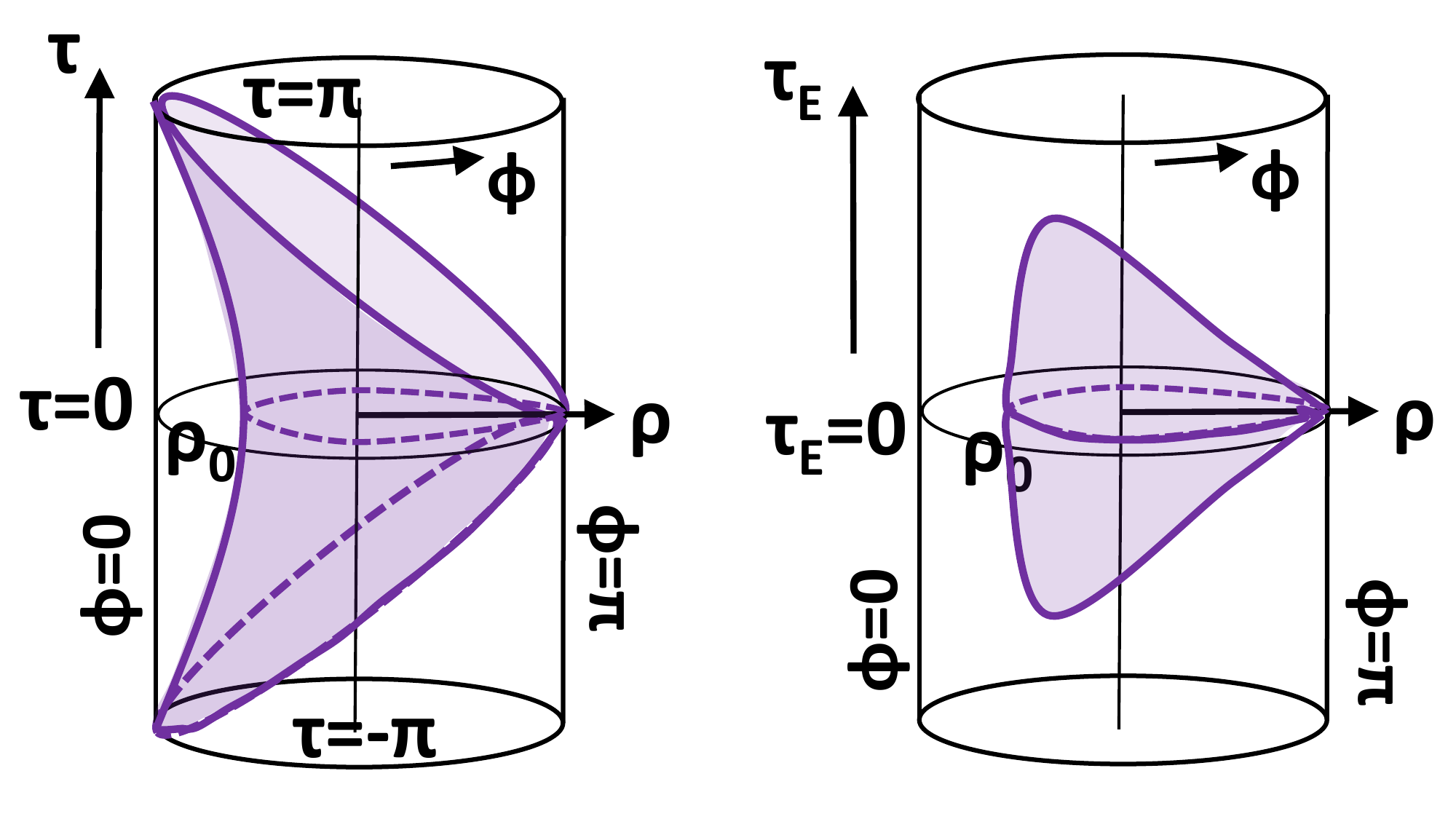}
		\caption{A sketch of the profile of a flat EOW brane in the Lorentzian global AdS (left) and in the Euclidean global AdS (right). We assume $R>z_0$.} 
		\label{fig:flatbrane}
\end{figure}

In the Poincare coordinate, the flat brane is very simply described such that $z$ takes a constant value: $z=z_0$, whose induced metric is obviously flat. 
In terms of the global coordinate, this reads 
\begin{align}
   \cosh{\rho}\cos{\tau}+\sinh{\rho}\cos{\phi}=\frac{R}{z_0}, \label{flatEOWB}
\end{align}
which is depicted in the left panel of Fig.\ref{fig:flatbrane}.
This EOW brane intersects with the global AdS boundary $\rho=\rho_\infty\to\infty$ along the diamond:
\ba
\cos\tau+\cos\phi=0, \label{bdyM}
\ea
which coincides with the boundary of Poincare AdS$_3$ as sketched in the right panel of Fig.\ref{fig:AdS}.
At $\tau=0$, it extends from the boundary point $(\rho,\phi)=(\infty,\pi)$ to the internal point $(\rho,\phi)=(\log\frac{R}{z_0},0)$ when $R>z_0$. When $R<z_0$, it extends to $(\rho,\phi)=(-\log\frac{R}{z_0},\pi)$.

In the Euclidean global AdS, by setting $\tau=i\tau_E$, the EOW brane is described by
\begin{align}
   \cosh{\rho}\cosh{\tau_E}+\sinh{\rho}\cos{\phi}=\frac{R}{z_0}.\label{EEOW}
\end{align}
This is depicted in the right panel of Fig.\ref{fig:flatbrane}.

\subsection{Three setups of flat brane holography}

\begin{figure}[ttt]
		\centering
		\includegraphics[width=10cm]{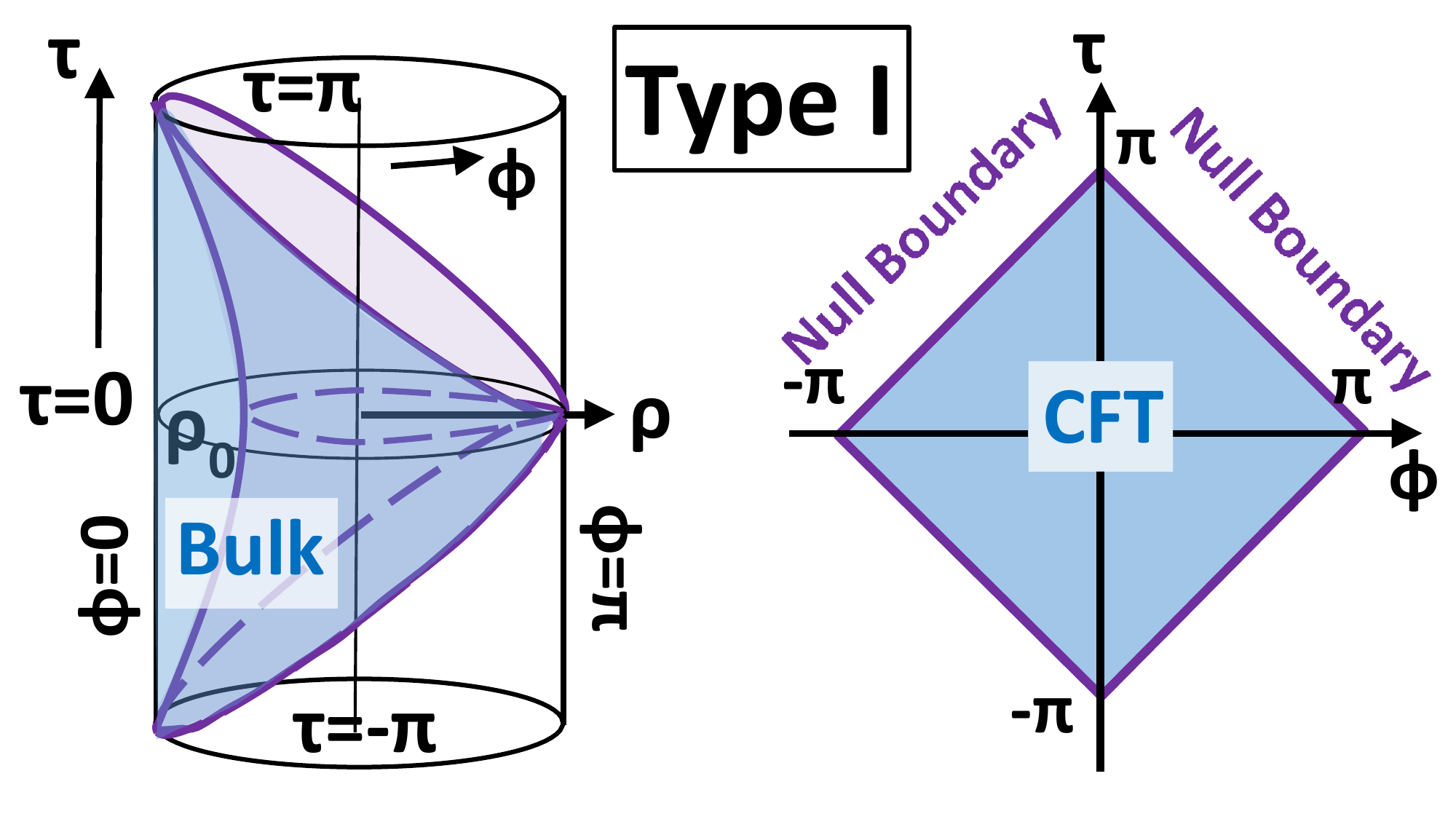}
        \includegraphics[width=4.7cm]{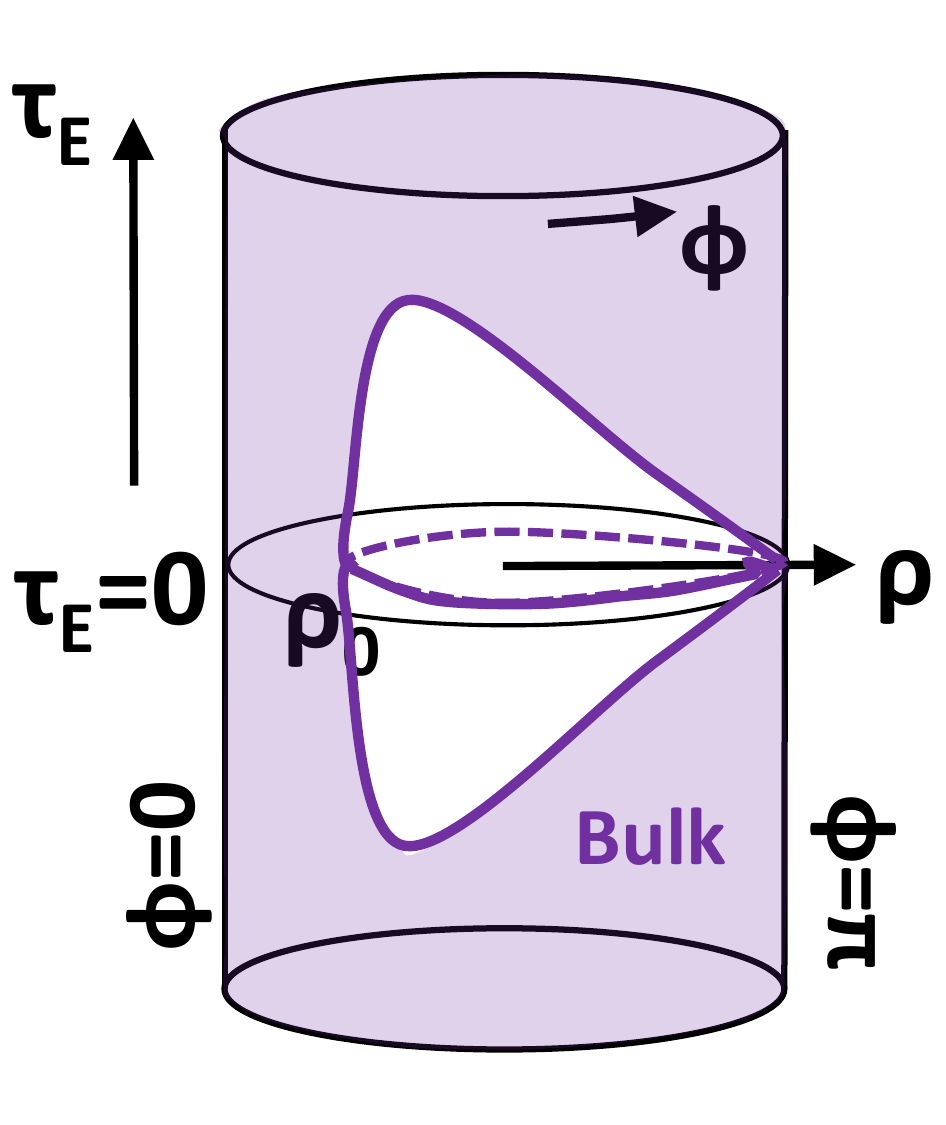}
        \includegraphics[width=10cm]{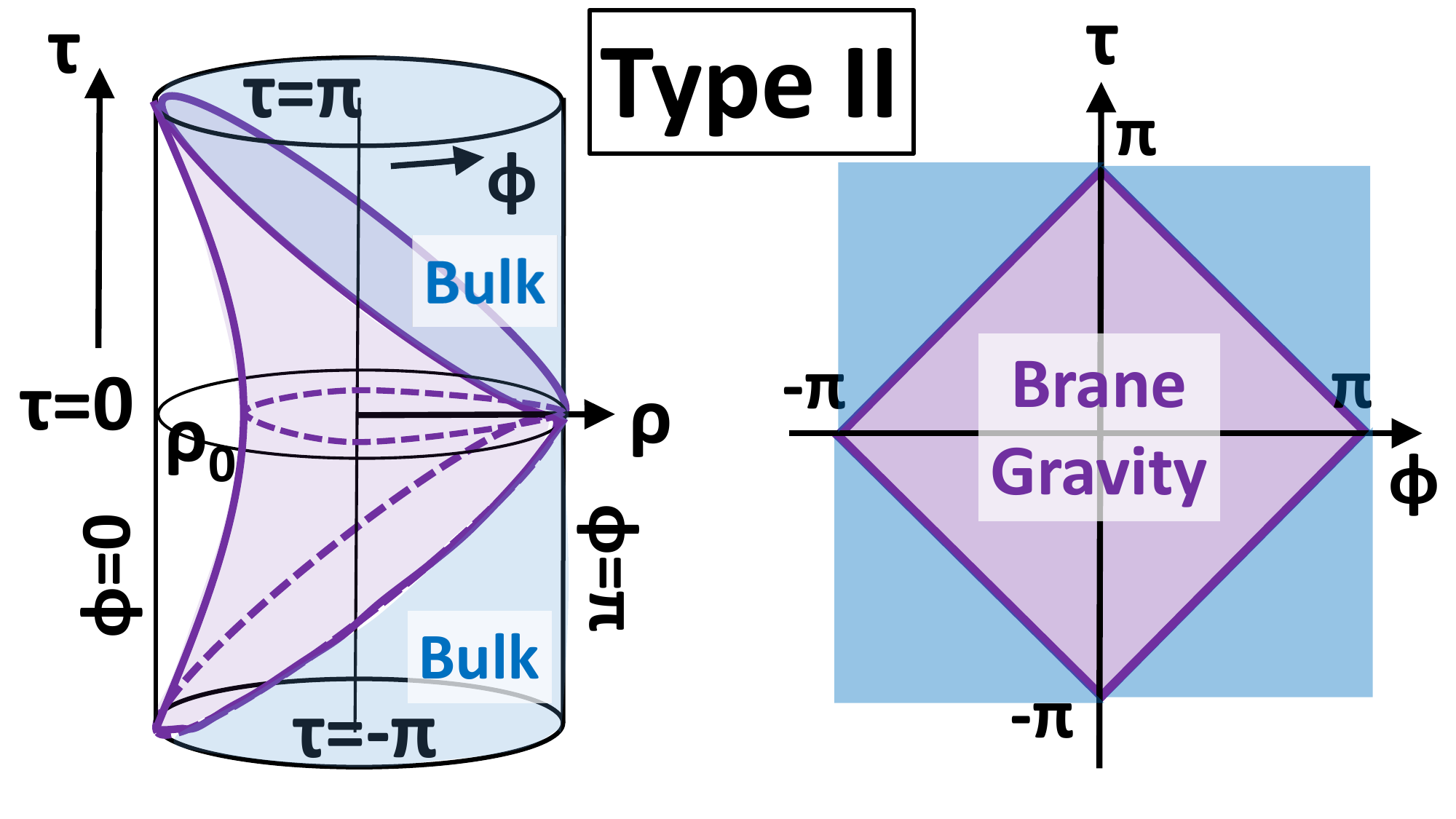}
         \includegraphics[width=4.7cm]{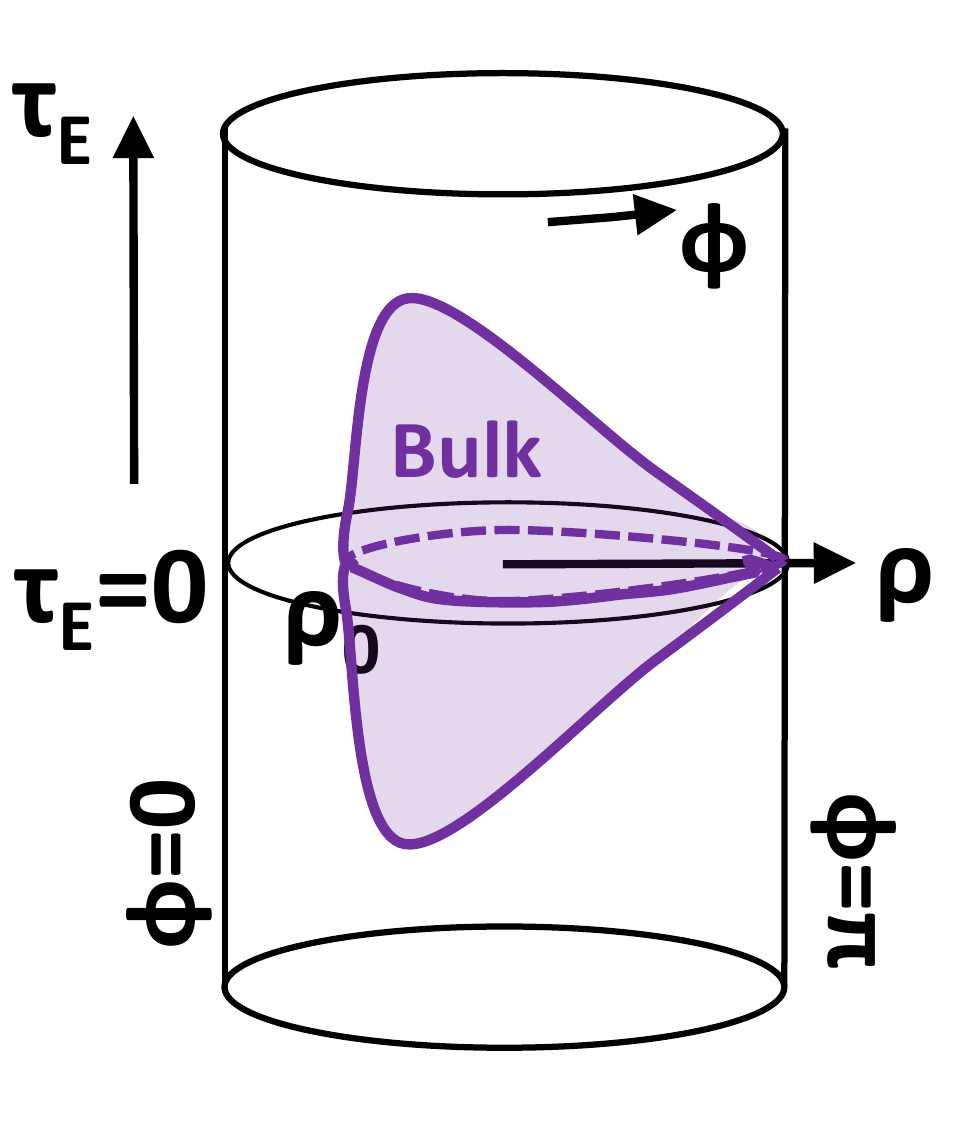}
        \includegraphics[width=10cm]{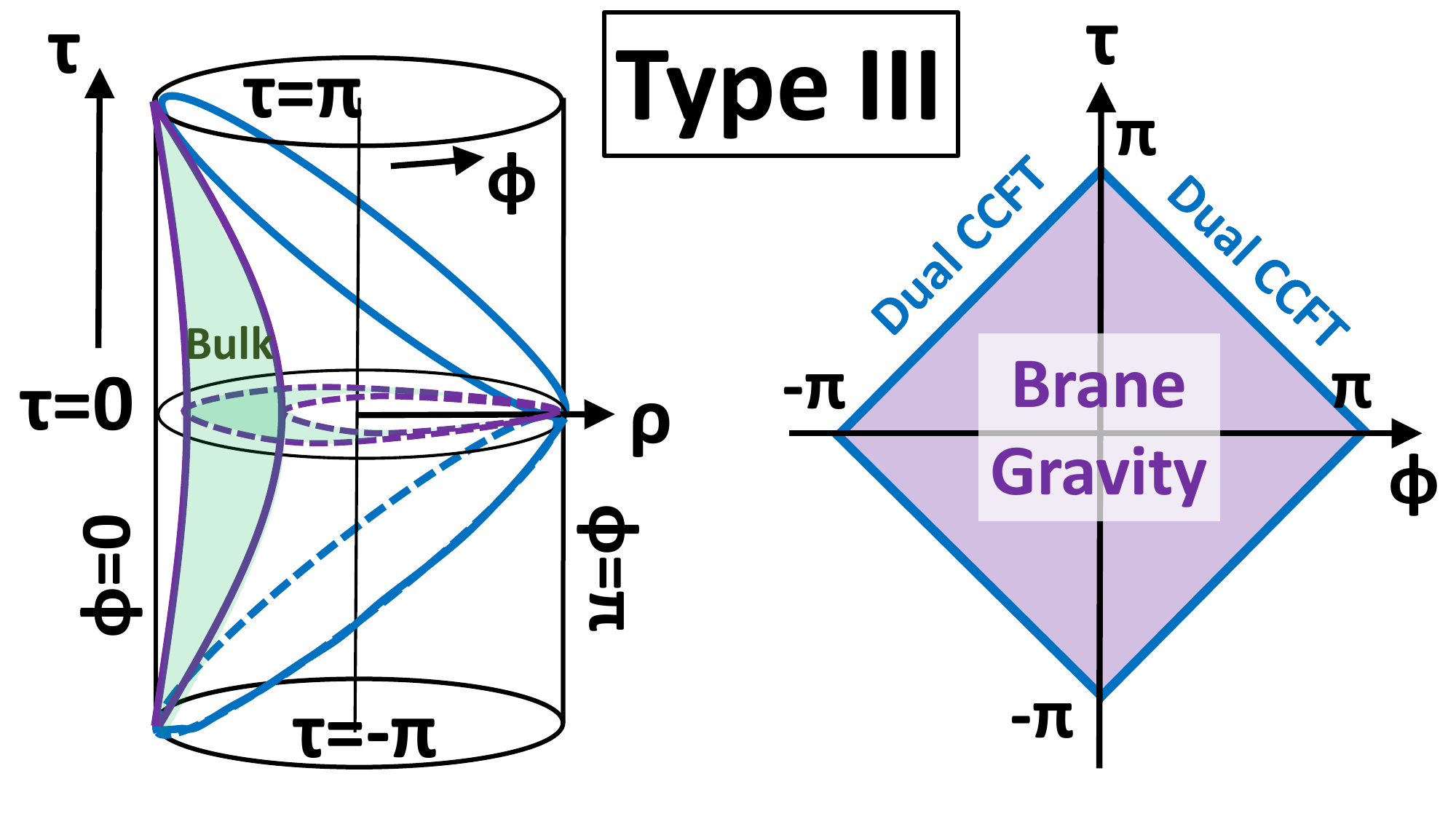}
         \includegraphics[width=4.7cm]{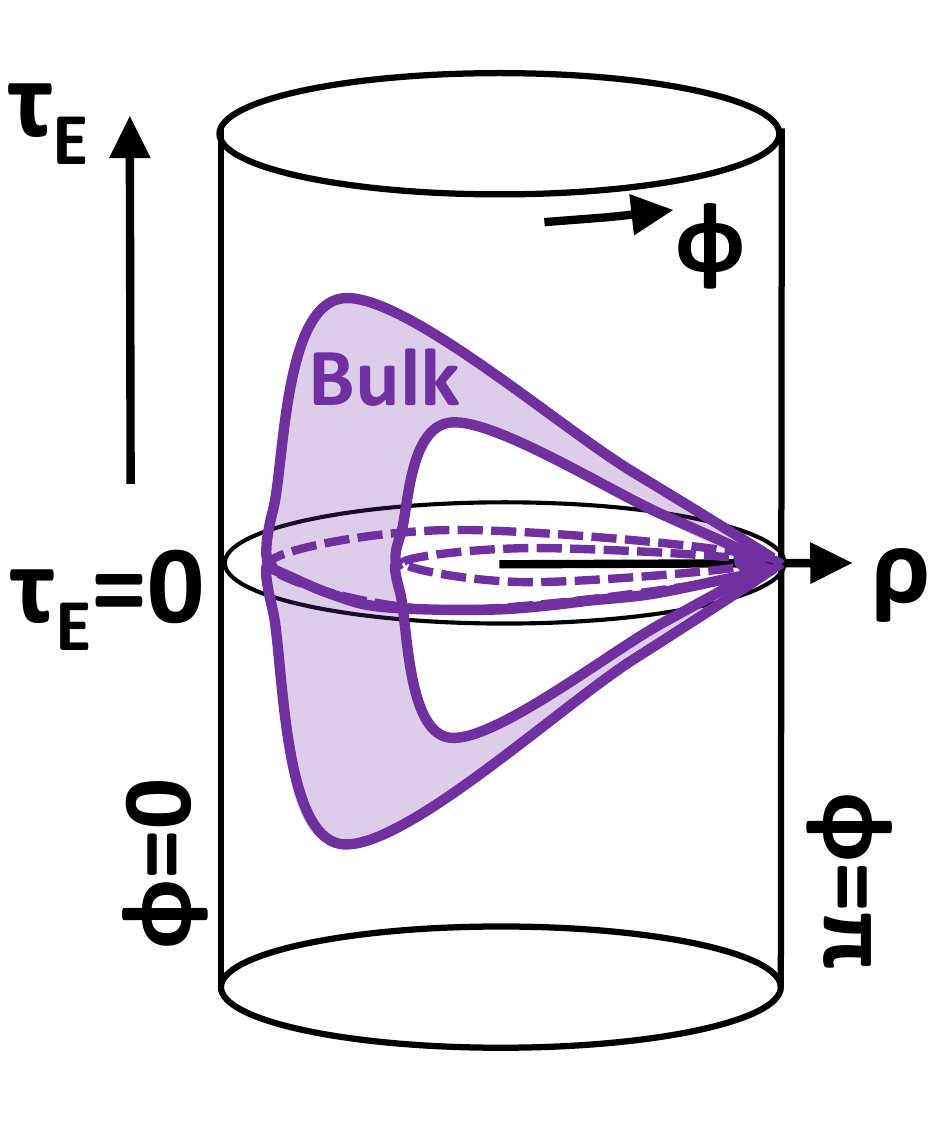}
		\caption{Sketches of the three setups: type I, II and III from the top to the bottom. The left panels show the regions of the gravity dual in the global AdS. The middle ones describe the dual two dimensional theories. The right ones show the Euclidean setups.} 
		\label{fig:setups}
\end{figure}

Using this flat EOW brane (\ref{flatEOWB}), we would like to consider the following three different setups inside the global AdS called 
type I, type II and type III. They are defined by 

\begin{align}\label{eq: types}
\begin{aligned}
 \mbox{Type I\,\,\,:} & \ \   0<z<z_0,\\
 \mbox{Type II\,\,:} & \ \   z>z_0,\\
 \mbox{Type III\,:} & \ \   z_1<z<z_2,
\end{aligned}
\end{align}
which are sketched in Fig.\ref{fig:setups}. Note that the gravity dual of type I includes the whole AdS boundary at $z=0$, which ends on the flat EOW brane at $z=z_0$. The gravity dual of type II is the complement of that of type I.  For the type III, we consider two EOW flat branes $z=z_1$ and $z=z_2$ and picks up the region between them. Below we will explain how the holography looks like. 

\subsection{Type I Setup}

The gravity dual in the type I case is given by the region between the AdS boundary and the flat EOW brane as sketched in the top left panel of Fig.\ref{fig:setups}. The AdS boundary is the diamond (times the celestial sphere $S^{d-2}$):
\ba
|\tau+\phi|\leq \pi,\ \ \ |\tau-\phi|\leq \pi, \label{diamond}
\ea
whose boundary is the solution to (\ref{bdyM}) as in the top middle panel of 
Fig.\ref{fig:setups}. Thus, its CFT dual is given by a CFT defined on the diamond (\ref{diamond}), with the null boundaries. The degrees of freedom of these null boundaries is expected to be equivalent to those of the flat EOW brane. From the observer in the gravity dual, this flat brane has the negative tension $T=-\frac{d-1}{R}$. 

The Euclidean setup of type I is shown in the top right panel of the same figure. This covers the whole region of the AdS boundary and thus its CFT dual describes a point-like defect in a CFT on $R\times S^{d-1}$. 

If we regard the EOW brane as a hard wall, the type I model looks similar to the AdS/QCD model with a hard wall e.g. 
\cite{Erlich:2005qh}. In this context the brane can be regarded as the IR cut off.

\subsection{Type II Setup}

Instead, in the type II setup, we pick up the interior region of the flat EOW brane with $T=\frac{d-1}{R}$, as depicted in the middle left panel of Fig.\ref{fig:setups}. 
This region can be infinitely extended in both the future and past direction $\tau\to\pm\infty$, which creates the AdS boundary region in these directions. Therefore, its CFT dual looks like the CFT which surrounds the diamond (\ref{diamond}) and its null boundaries. Via the double holography, this is equivalent to the CFT coupled to the flat space gravity on the diamond as depicted in the middle center panel of Fig.\ref{fig:setups}. Hence, this provides an interesting model which includes the flat space holography. Note also that this type II setup in the Poincare patch is 
equivalent to the RS2 model \cite{Randall:1999vf}, though in this paper we consider its global extension in order to have the asymptocially AdS UV boundary where the dual CFT lives. 

 The Euclidean setup of type II is shown in the  middle right panel of the same figure. Its AdS boundary includes only the point $\phi=\pi$. Therefore, the gravity in the type II region is expected to dual to a point-like theory.

\subsection{Type III Setup}
\label{subsec: III setup}
In the type II setup, the field theory dual includes the large `bath' theory, which extends outside of the diamond. To extract the flat space holography, we can remove the large bath by adding one more flat EOW brane and by focusing on the region between the two EOW branes, sketched in the bottom panels of Fig.\ref{fig:setups}. This consideration leads to the type III setup.
The AdS boundary of this model only consists of the null edges of the diamond. Thus, its CFT dual is given by a $d-1$-dimensional Carollian CFT (CCFT$_{d-1}$) which lives on the null lines $|\tau+\phi|=|\tau-\phi|=\pi$. Note that the insertion of the flat EOW brane breaks the isometry of the AdS$_{d+1}$ with the remaining symmetry group $ISO(1,d-1)$, which is also the global symmetry group of the CCFT$_{d-1}$. Via the double holography, we can also regard this as the flat space gravity on the diamond. 
Thus, this provides an interesting setup of flat space holography.  Note also that this type III setup in the Poincare patch is equivalent to the RS1 model \cite{Randall:1999ee}, though in this paper we consider its global extension in order to have the asymptocially AdS boundary which is codimension two. 

In the Euclidean setup of type III, there is also the region surrounded by the two EOW branes, and its AdS boundary only includes the point $\phi=\pi$, which shows that the gravity in the type III region should again be dual to a point-like theory.


\section{Holographic entanglement entropy}
\label{sec:HEE}
In this section, we calculate the entanglement entropy of the subsystem in the dual field theory holographically in the AdS$_3$ in type I, II setups and in AdS$_4$ in type III setup by holographic entanglement entropy formula \cite{Ryu:2006bv,Ryu:2006ef,Hubeny:2007xt}. 

In the type I, II setups, given the presence of an EOW brane in the bulk AdS$_3$ spacetime, two types of extremal surfaces arise according to the AdS/BCFT correspondence. The first type is the connected extremal surface, denoted by $\Gamma^{\con}_{\mathcal{A}}$, with its boundary located solely on the entangling surface $\partial \mathcal{A}$. The second type is the disconnected surface $\Gamma^{\dis}_{\mathcal{A}}$, anchored on the EOW brane and also $\partial\mathcal{A}$. Specifically, $\Gamma_{A}^{\dis}$ represents geodesics that originate from one of the entangling surfaces and terminate on the EOW brane. The endpoint of $\Gamma_{A}^{\dis}$ on the EOW is determined by the extremization conditions. The contributions to the entanglement entropy are denoted as $S_{\mathcal{A}}^{\mathrm{con}}$ and $S_{\mathcal{A}}^{\mathrm{dis}}$ for the connected and disconnected surfaces, respectively. By applying the holographic entanglement entropy (HEE) formula, we can determine the entanglement entropy of the bath,
\begin{equation}
S_{\mathcal{A}} = \min \{S_{\mathcal{A}} ^{\mathrm{con}}, S_{\mathcal{A}} ^{\mathrm{dis}}\} = \min \left\{ \frac{\text{Area}(\Gamma_{\mathcal{A}}^{\rm{con}})}{4\GN}, \frac{\text{Area}(\Gamma_{\mathcal{A}}^{\rm{dis}})}{4\GN}\right\} \,,  \label{minchoice}
\end{equation}
where $\GN$ denotes the Newton's constant for the gravitational theory in the AdS$_3$ bulk spacetime. We set the AdS radius to be unit $R=1$ and the central charge of the dual two-dimensional CFT is given by $c=\frac{3}{2G_N}$.
In the AdS$_3$ case, the area of the extremal surface becomes the geodesic length which can be derived from the chordal distance,
\begin{equation}\label{eq:geo length from chordal}
   \cosh D_{AdS}=-X(1)\cdot X(2),
\end{equation}
where the AdS radius $R$ is taken to be $1$, $X(i)$ represents the coordinate of the $i$-th point in the embedding coordinate and the inner product is calculated by the metric of the embedding spacetime.

\begin{figure}[ttt]
		\centering
		\includegraphics[width=10cm]{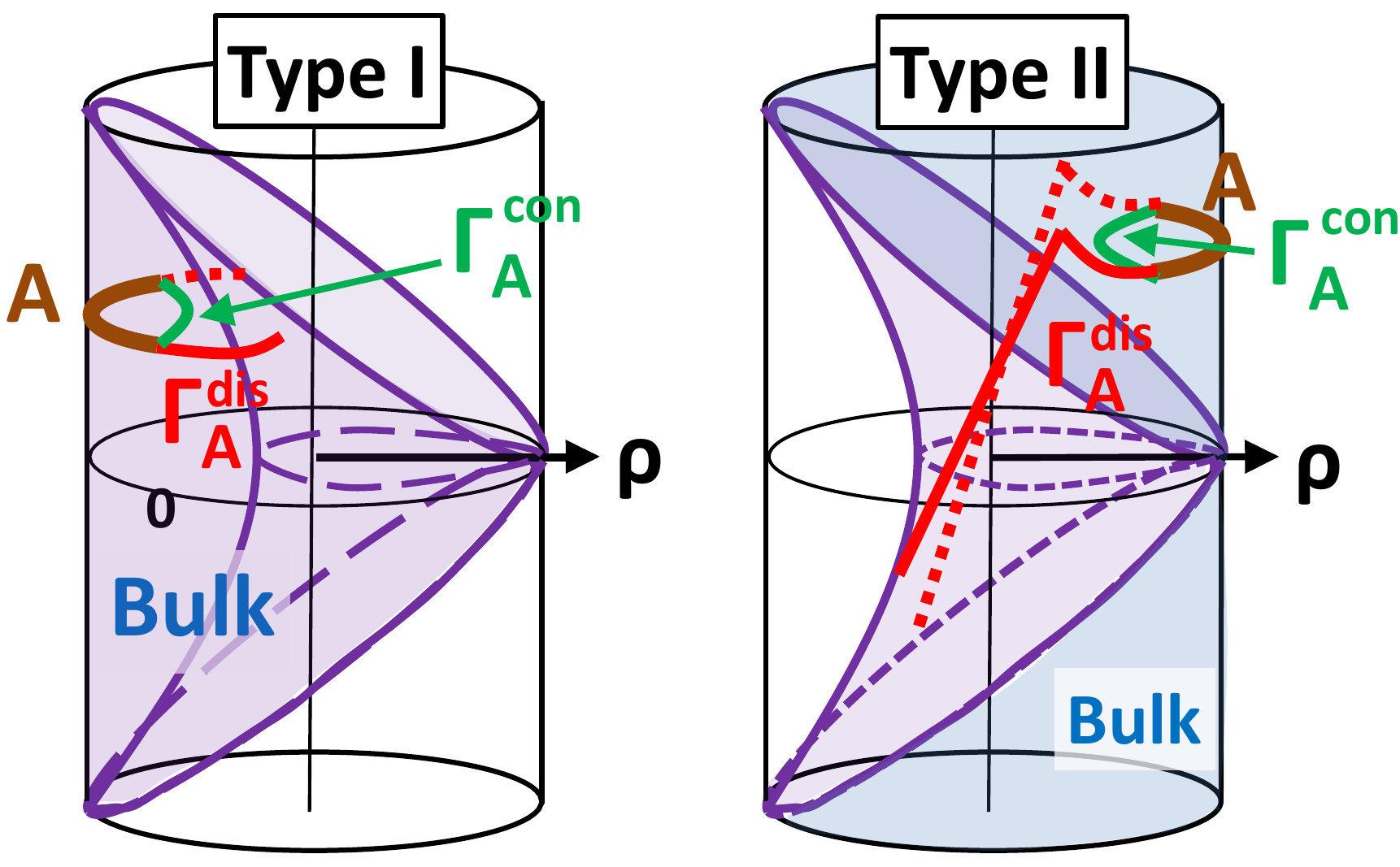}
		\caption{A sketch of calculations of holographic entanglement entropy in type I (left) and type II (right) setup for AdS$_3$. The connected and disconnected geodesic are described by the gree and red curves.} 
		\label{fig:HEE}
\end{figure}

We examine both connected and disconnected extremal surfaces and determine their lengths in type I and type II configurations, as detailed in subsections \eqref{subsec: hee I} and \eqref{subsec: hee II}, and illustrated in Fig. \ref{fig:HEE}.

\subsection{Holographic entanglement entropy in type I setup}
\label{subsec: hee I}
It is convenient to consider the type I setup in the Poincare AdS coordinate,
\begin{equation}
    ds^2=\frac{1}{z^2}(-dt^2+dz^2+dx^2),
\end{equation}
so that the length is
\begin{equation}
    \cosh D_{Poin}=\frac{x_{12}^2-t_{12}^2+z_1^2+z_2^2}{2z_1z_2},
\end{equation}
for the geodesic connecting two points $(z_1,t_1,x_1)$ and $(z_2,t_2,x_2)$ in AdS$_3$, which can be derived from \eqref{eq:geo length from chordal} and the embedding \eqref{eq: embedding} for the Poincare coordinate, where $x_{ij}=x_i-x_j,~t_{ij}=t_i-t_j$. The interval $\mathcal{A}$ is labeled by its two endpoints on the asymptotic boundary of the AdS on $z=0$,
\begin{equation}
    \partial\mathcal{A}=\{A_1=(z=0,t_1,x_1),~A_2=(z=0,t_2,x_2)\}.
\end{equation}
\paragraph{Connected phase contribution} The holographic entanglement entropy from the connected phase is given by the length of the geodesic connecting the two endpoints with proper regulator on $z=z_c\to0$,
\begin{equation}
    S_{I}^\con = \frac{ D_{12} }{4 G_N}=\frac{1}{4G_N}\qty(\log(x_{12}^2-t_{12}^2)-\log z_{c1}-\log z_{c2}).
\end{equation}
In the original global coordinate, this is written as
\begin{equation}
    S_{I}^\con = \frac{ D_{12} }{4 G_N}=\frac{1}{4G_N}\qty(\log\frac{\cos\tau_{12}-\cos\phi_{12}}{2}-2\log\epsilon ),  \label{eqconhe}
\end{equation}
which is the same as the entanglement entropy for the interval on the cylinder in CFT$_2$. Here we write the UV cut off in the global coordinate  $\ep=e^{-\rho_\inf}$ in terms of those in the Poincare AdS $z_{c1}$ and $z_{c2}$ via $z_{c1}=\frac{2}{\cos\tau_{1}+\cos\phi_{1}}\ep$ and similar for $z_{c2}$ as follows from \eqref{eq: embedding}.

\paragraph{Disconnected phase contribution} To get the contribution from the disconnected phase, we should consider the geodesic connecting one of the endpoints of the interval, e.g., $A_1$ without loss of generality, and the point $A_3(z_0,t_3,x_3)$ on the EOW brane $z=z_0$, whose length is given by
\begin{equation}
    D_{13}= \log(\frac{x_{13}^2-t_{13}^2+z_0^2}{z_0})-\log z_{c1},
\end{equation}
where we have considered the regulator on $z_{c1}\to0$. The holographic entanglement entropy from the disconnected phase is determined by the extremal condition
\begin{equation}
S^{\text{dis}}_{\mathcal{A}}=\underset{A_3}{\operatorname{Ext}}\frac{\text{Area}(D_{13})}{4G_N}+\underset{A_4}{\operatorname{Ext}}\frac{\text{Area}(D_{24})}{4G_N},
\end{equation}
where the counter-partner of the geodesic connecting the endpoint $A_2$ and the point $A_4$ on the brane is also taken into consideration. This, in turn, gives the location of $A_3$,
\begin{equation}
    \partial_{t_3}D_{13}=0,\ \ \partial_{x_3}D_{13}=0.
\end{equation}
It yields the solution as
\begin{equation}\label{eq: sol dis I}
    t_3=t_1,\ \ x_3=x_1.
\end{equation}
A further calculation of the second derivatives of the geodesic length shows that
\begin{equation}
    \partial_{t_3}^2D_{13}<0,\ \ \partial_{x_3}^2D_{13}>0,
\end{equation}
which shows the disconnected extremal surface extremized along the flat brane is locally maximal with respect to temporal variations and minimal with respect to spatial variations. Consequently, it is a proper candidate surface for the holographic entanglement entropy, as it is reasonable to infer that this type of extremal surface corresponds to the dominant saddle in the Euclidean path integral for calculating entanglement entropy.

With the solution \eqref{eq: sol dis I}, we find the contribution to the holographic entanglement entropy from the length of the geodesic in disconnected phase
\begin{equation}
    S_{I}^{\text{dis}}=2\frac{D_{dis}}{4G_N}=2\log\frac{z_0}{z_{c1}},
\end{equation}
where we have considered the contribution from the other geodesic connecting $A_2$ and the point on the flat EOW brane, resulting in an additional identical contribution. 

Thus we can determine the disconnected holographic entanglement entropy in the $(\tau,\phi)$ coordinate system as follows
\begin{equation}\label{eq: dis I global length}
S_{I,cyl}^{\text{dis}}=\frac{D_{dis,13}}{4G_N}+\frac{D_{dis,24}}{4G_N}=\frac{1}{4G_N}\qty(\log\frac{z_0(\cos\tau_1+\cos\phi_1)}{2\epsilon}+\log\frac{z_0(\cos\tau_2+\cos\phi_2)}{2\epsilon}),
\end{equation}
where $\ep=e^{-\rho_\inf}$ is the UV cut off in the global AdS$_3$, which is related to $z_c$ via $z_c=\frac{2}{\cos\tau_{1,2}+\cos\phi_{1,2}}\ep$ as \eqref{eq: embedding}.
The actual holographic entanglement entropy is given by the minimum among the connected entropy (\ref{eqconhe}) and disconnected one (\ref{eq: dis I global length}), by applying the general rule (\ref{minchoice}).

\subsection{Holographic entanglement entropy in type II setup}
\label{subsec: hee II}
Compared to the type I setup, the situation is significantly altered in the type II setup, where the dual field theory with large bath extends out of the Poincare patch. We then consider the global AdS$_3$ coordinate,
\begin{equation}
    ds^2=-\cosh^2\rho d\tau^2+d\rho^2+\sinh^2\rho d\phi^2.
\end{equation}
The corresponding field theory covers the region
\begin{equation}\label{eq: II region}
    \cos\phi+\cos\tau<0,
\end{equation}
in the asymptotic boundary $\rho\to\infty$ of the AdS$_3$ spacetime. The flat EOW brane is discribed by \eqref{flatEOWB}. In this global coordinate, the geodesic length is represented by
\begin{equation}
    \cosh D_{global}=\cos(\tau_1-\tau_2)\cosh\rho_1\cosh\rho_2-\cos(\phi_1-\phi_2)\sinh\rho_1\sinh\rho_2,
\end{equation}
for the geodesic connecting two points $(\rho_1,\tau_1,\phi_1)$ and $(\rho_2,\tau_2,\phi_2)$. We specify the interval $\mathcal{A}$ by the two endpoints,
\begin{equation}
    \partial\mathcal{A}=\{A_1=(\rho=\rho_\inf\to\inf,\t_1,\p_1),~A_2=(\rho=\rho_\inf\to\inf,\t_2,\p_2)\}.
\end{equation}
\paragraph{Connected phase contribution} The holographic entanglement entropy from the connected phase is given by the length of the geodesic connecting the two endpoints $A_1,A_2$ and then expand with respect to the regulator $\rho_\inf=-\log\epsilon$ at $\epsilon=0$,
\begin{equation}
    S_{II}^\con = \frac{ D_{12} }{4 G_N}=\frac{1}{4G_N}\qty(\log\frac{\cos\tau_{12}-\cos\phi_{12}}{2}-2\log\epsilon ),\label{conII}
\end{equation}
which is identical to (\ref{eqconhe}).

\paragraph{Disconnected phase contribution} To obtain the contribution from the disconnected phase, we consider the geodesic connecting one endpoint of the interval $A_1$ and the point $A_3(\rho_3,\t_3,\p_3)$ on the EOW brane \eqref{flatEOWB}. The length of this geodesic is given by
\begin{equation}
    D_{13}= \log\big(\cos(\tau_1-\tau_3)\cosh\rho_3-\cos(\phi_1-\phi_3)\sinh\rho_3\big)-\log \epsilon,
\end{equation}
where we have expanded the result with recpect to $\epsilon$. Similarlly, the location of the point $A_3$ on the flat EOW brane is determined by the extremal condition
\begin{equation}
    \partial_{\tau_3} D_{13}=0,\ \ \partial_{\rho_3} D_{13}=0,
\end{equation}
if the coordinate $\phi_3$ is considered as a function of $\tau_3$ and $\rho_3$, as determined by \eqref{flatEOWB}. Alternatively, any two of the coordinates among $(\rho_3, \tau_3, \phi_3)$ can be used to express the extremal condition, with the remaining coordinate regarded as a function of these two. However, any of them is hard to solve since they are transcendental equations.

To proceed, we consider the coordinate transformation of $(\r_3,\t_3,\p_3)$ to the Poincare coordinate. Or equivalently, one global coordinate and one Poincare coordinate in the embedding can be replaced into \eqref{eq:geo length from chordal}, resulting
\begin{equation}\label{eq: geo glo poin}
    D_{13}= \log\frac{\cos \tau_1\left(1-t_3^2+x_3^2+z_0^2\right)+\cos
   \phi_1 \left(-1-t_3^2+x_3^2+z_0^2\right)+2  t_3 \sin\tau_1-2 x_3 \sin \phi_1}{2z_0}-\log\e,
\end{equation}
where the point $A_3$ on the flat EOW $z=z_0$ is represented in the Poincare coordinate while the endpoint $A_1$ is represented in the global coordinate. In this case, the extremal condition to determine the location of $A_3$ is given by
\begin{equation}
    \partial_{t_3}D_{13}=0,\ \ \partial_{x_3}D_{13}=0.
\end{equation}
The solution is
\begin{equation}\label{eq: II saddle}
    t_3=\frac{ \sin \tau_1}{\cos \tau_1+\cos \phi_1},\ \ \ x_3=\frac{ \sin \phi_1}{\cos \tau_1+\cos \phi_1}.
\end{equation}
Similarly, we consider the second derivatives on the geodesic length with respect to the temporal and spatial directions
\begin{equation}
    \partial_{t_3}^2D_{13}<0,\ \ \partial_{x_3}^2D_{13}>0,
\end{equation}
indicating that in the type II setup, this is also a good candidate for the disconnected geodesic since it is locally maximal with respect to temporal variations and minimal with respect to spatial variations. Then with the solution \eqref{eq: II saddle}, we find the contribution to the holographic entanglement entropy from the length of the geodesic in disconnected phase
\begin{align}\label{eq: dis II length}
S_{II}^{\text{dis}}&=\frac{D_{dis,13}}{4G_N}+\frac{D_{dis,24}}{4G_N}\nonumber\\
&=\frac{1}{4G_N}\qty(\log\frac{z_0(\cos\tau_1+\cos\phi_1)}{2\epsilon}+\log\frac{z_0(\cos\tau_2+\cos\phi_2)}{2\epsilon}),
\end{align}
where the extra contribution from the other geodesic connecting $A_2$ and the point on the flat EOW brane is also added. This expression of the geodesic length is the same as that in the type I case but the difference is the range of $\cos\tau_{1,2}+\cos\phi_{12}$ which is positive in the type I case and negative in the type II case.
Meanwhile, we should choose the same positive sign of $z_0$ in both cases so that compared to the type I setup where the disconnected geodesic is space-like, in the type II setup, the disconnected geodesic contains both time-like and space-like pieces. In summary, the disconnected geodesic in type II  is found to be as follows:
\begin{equation}\label{eq: dis II lengthcompa}
    S_{II}^{\text{dis}}=\frac{1}{4G_N}\qty(\log\frac{z_0|\cos\tau_1+\cos\phi_1|}{2\epsilon}+\log\frac{z_0|\cos\tau_2+\cos\phi_2|}{2\epsilon})
    +\frac{\pi}{2G_N}i.
\end{equation}
Since this is complex valued, this should be interpreted as the pseudo entropy \cite{Nakata:2021ubr}. We argue that the final holographic (pseudo) entanglement entropy is given by selecting 
either $S_{II}^\con$ (\ref{conII}) or $ S_{II}^{\text{dis}}$ in (\ref{eq: dis II lengthcompa}) such that the real part is smaller. 

Since the imaginary part, which comes from the time-like geodesic as discussed soon below, takes the form 
$\frac{\pi i}{3}c$, this may be regarded as a version of time-like entanglement entropy \cite{Doi:2022iyj,Doi:2023zaf}. 
The presence of this imaginary part show that the dual density matrix (or more properly called transition matrix) is not hermitian. Thus it is possible that the dual theory is non-unitary. In particular, since the AdS itself should be dual to a unitary CFT, this exotic feature should be due to the flat EOW brane. This suggests that the flat space gravity localized on the EOW brane is dual to a non-unitary field theory, in spite that the gravity theory in flat space is unitary, which was also implied from the fact that central charge of the dual celestial CFT is imaginary \cite{Ogawa:2022fhy}. 

\paragraph{Possible profiles of Disconnected geodesic} Let us study an explicit profile of the disconnected geodesic which connects the endpoint $A_1(\tau_1,\rho_\infty,\phi_1)$ of the interval and the point $A_3(\tau_3,\rho_3,\phi_3)$ (determined by \eqref{eq: II saddle} and the further coordinate transformation to the global coordinate) on the flat EOW brane. Firstly one considers the point $A_3'(\tau_3+\pi,\rho_3,\phi_3+\pi)$ on the flat EOW brane $z=-z_0$. Then one can find an arbitrary point $B(\tau_B,\rho_B,\phi_B)$ on the geodesic connecting $A_1$ and $A_3'$ which is space-like. There is a corresponding point $B'(\tau_B+\pi,\rho_B,\phi_B+\pi)$ on the space-like geodesic connecting $A_1'(\tau_1+\pi,\rho_\inf,\phi_1+\pi)$ and $A_3$.

The total disconnected geodesic is then contains the following pieces: the space-like geodesic connecting $A_1$ and $B$, the time-like geodesic connecting $B$ and $B'$, and also the space-like geodesic connecting $B'$ and $A_3$. We sketched in Fig.\ref{fig:HEE} for the case where $B$ coincides with $A'_3$ as an example. 

The length of the time-like piece is
\begin{equation}
    \cosh D_{BB'}=-1,\ \ D_{BB'}=i\pi.
\end{equation}
The total length of the space-like pieces is
\begin{equation}
D_{A_1B}+D_{B'A_3}=D_{A_1A_3'}=D_{A_1'A_3}=\log\frac{-z_0(\cos\tau_1+\cos\phi_1)}{2\epsilon}.
\end{equation}
The total contribution reproduce \eqref{eq: dis II length}. Among the various choices of the point $B$, we can choose $B=A_3'$ especially so that the total disconnected geodesic intersects with the flat EOW brane $z=z_0$ only once. This is depicted in the right panel of Fig.\ref{fig:HEE}.

Though the geodesic we found in this above, gives the correct geodesic length based on which we computed the holographic entanglement entropy (\ref{eq: dis II lengthcompa}), we have to admit that it looks like an adhoc prescription. We would like to note that a similar situation is known for the time-like entanglement entropy \cite{Doi:2022iyj,Doi:2023zaf}, which is also given in terms of a union of time-like and space-like geodesic.  It is possible that we can also have an interpretation in terms of complex geodesics as in \cite{Heller:2024whi, Nunez:2025ppd, Heller:2025kvp}.
We would like to leave a further study of its interpretation for a future work.

\subsection{Holographic entanglement entropy in type III setup}
\label{subsec: hee III}
We will then consider the holographic entanglement entropy in type III setup with the co-dimension two wedge holography. As introduced in section \eqref{subsec: III setup}, there exist two flat EOW branes and the dual field theory is the CCFT$_{d-1}$ on the null surface $|\tau+\phi|=|\tau-\phi|=\pi$ in AdS$_{d+1}$. The entanglement entropy is well-studied in the two dimensional CCFTs so we will consider the AdS$_4$ bulk ($d=3$) in this subsection.

\begin{figure}[ttt]
		\centering
		\includegraphics[width=6cm]{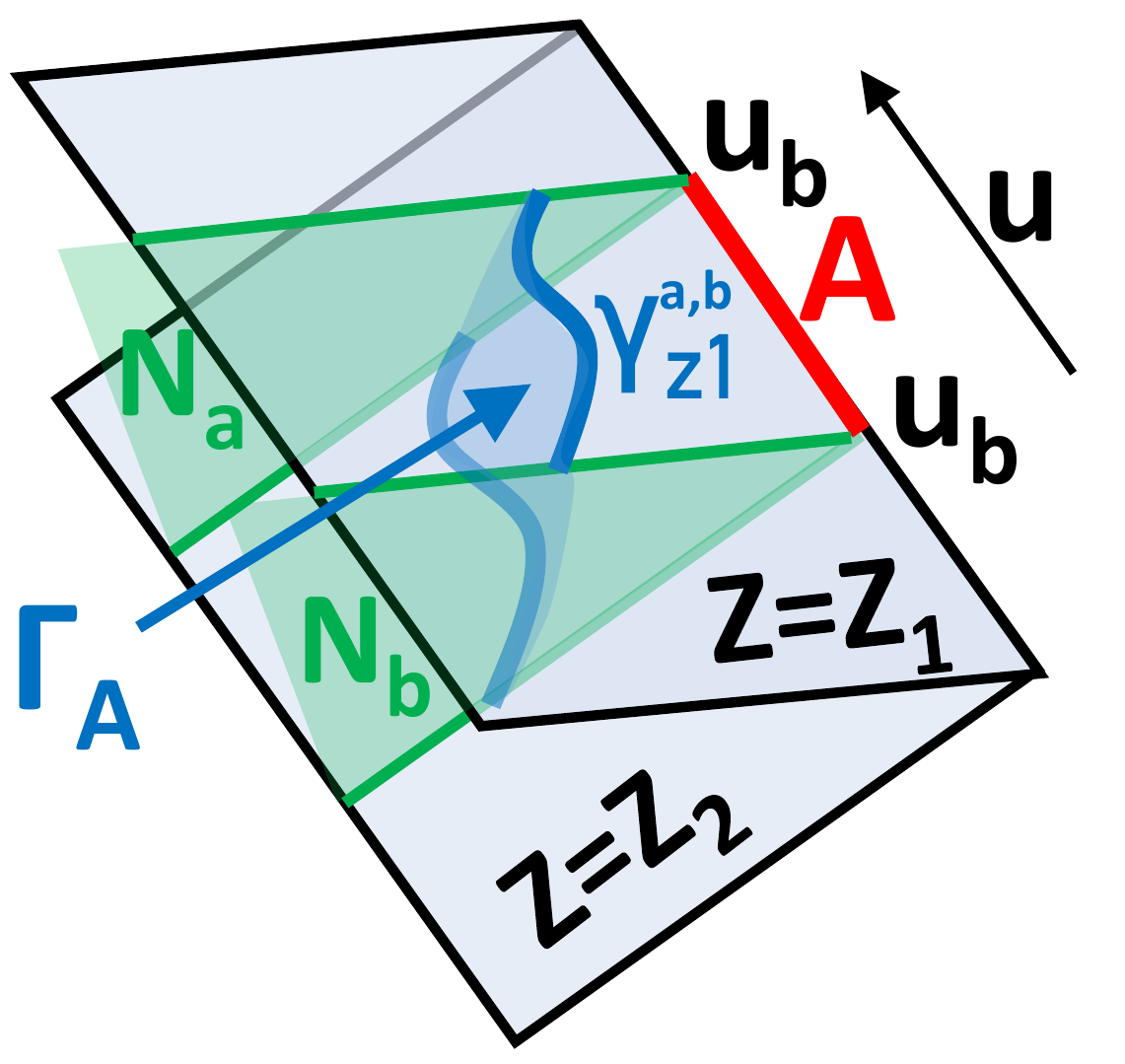}
		\caption{A sketch of calculations of holographic entanglement entropy in type III setup. The two light blue colored planes are EOW branes at $z=z_1$ and $z=z_2$, respectively, which intersects at the global AdS boundary, where we choose the subsystem $A$. The extremal surface which computes the entanglement entropy is $\Gamma_A$ which stretches between two null surfaces $N_a$ and $N_b$.
        This reproduces the swingle surface calculation in the flat space holography for the CCFT.} 
		\label{fig:Swing}
\end{figure}

We consider AdS$_4$ in Poincare coordinate
\begin{equation}\label{eq: ads4 poin}
    ds^2=\frac{1}{z^2}(dz^2+ds_b^2),\ \ ds_b^2=-dt^2+dx^2+dy^2=-dt^2+dr^2+r^2d\Phi^2=-du^2-2dudr+r^2d\Phi^2,
\end{equation}
where the last equations come from a further transformation to the coordinate $(z,t,r,\Phi)$
\begin{equation}\label{eq: flat trans}
t=u+r,\ \ x=r\cos\Phi,\ \ y=r\sin\Phi.
\end{equation}
The point in the null boundary of the constant $z$ slice as $r\to\infty$ becomes the point 
\begin{equation}
    \rho\to\infty,\ \ \cos^2\tau=\cos^2\phi=\frac{u^2}{1+u^2},\ \ \tan\theta=\tan \Phi,
\end{equation}
in the global coordinate $(\rho,\tau,\phi,\theta)$
\begin{equation}
    ds^2=-\cosh^2{\rho}d\tau^2+d\rho^2+\sinh^2{\rho}(d\phi^2+\sin^2\phi d\theta^2).
\end{equation}
Note that it is $z$-independent so that the null boundary of the slices with different $z$ corresponds to the same null surface $\cos^2\tau=\cos^2\phi$, where the CCFT lives. With two flat EOW branes at $z=z_1,z_2$ inserted, the intersection is the above null surface. The interval $A$ is considered also in this null surface, labeled by its two endpoints $A_a(u_a,\Phi_a)$ and $A_b(u_b,\Phi_b)$. We argue that the holographic entanglement entropy for the interval $A$ is
\begin{equation}
    S_A=\frac{\text{Area}(\Gamma_A)}{4G_N},
\end{equation}
where $\Gamma_A$ is the two-dimension surface determined in the following steps:
\begin{itemize}
    \item[1] Find the null geodesics $\gamma_{z}^{a,b}$ on the constant $z$ slices, with $\partial \gamma_{z}^{a,b}=A_{a,b}$. The tangent vector of the null geodesics are the bulk modular flow generators on fixed $z$ slices between the two flat EOW branes $z_1<z<z_2$. Then they determines two null surfaces $N_a,N_b$ which are the unions of $\gamma_{z}^{a,b}$ with $z_1<z<z_2$.
    \item[2] Consider two arbitrary curves $\Gamma_a,\Gamma_b$ on the null surface $N_a,N_b$, with the two endpoints on the two flat EOW brane $z=z_{1,2}$, parametrized by $r_{a,b}(z)$.
    \item[3] Find the two-dimension extremal surface $\Gamma_{A}$ in the bulk region bounded by the two branes, with $\partial \Gamma_A= \Gamma_a\cup \Gamma_b$.
    \item[4] Minimize the area of $\Gamma_A$ with respect to the two curves $\Gamma_{a,b}$ on the two null surfaces $N_{a,b}$.
    \item[5] The final surface which computes the entanglement entropy is given by the combination of part of $N_{a,b}$ and $\Gamma_{A}$. The range of $N_{a,b}$ is bounded by the curves $\Gamma_{a,b}$ found in the step 4. Since $N_{a,b}$ are null, without contribution to the area, the holographic entanglement entropy is captured by the surface $\Gamma_A$.
\end{itemize}
This is a generalization of the holographic entanglement entropy in the wedge holography \cite{Akal:2020wfl} for the AdS branes to our flat branes. This fits nicely with the swing surface proposal in Flat$_3$/CCFT$_2$ correspondence \cite{Jiang:2017ecm,Apolo:2020bld}. As in the swing surface proposal, the bench is described by the parametric equations
\begin{align}\label{eq: bench equation}
\begin{aligned}
x&=\csc (\Phi_a-\Phi_b) \qty((u_a-t) \sin \Phi_b+(t-u_b) \sin \Phi_a),\\
y&=-\big(\csc (\Phi_a-\Phi_b) \qty((u_a-t) \cos \Phi_b+(t-u_b) \cos \Phi_a)\big),
\end{aligned}
\end{align}
and the range of $t$ is
\begin{equation}\label{eq: t range}
    t\in\qty[\frac{u_a-u_b}{\cos (\Phi_a-\Phi_b)-1}+u_a,\frac{u_b-u_a}{\cos (\Phi_a-\Phi_b)-1}+u_b].
\end{equation}
We will verify later that \eqref{eq: bench equation} and \eqref{eq: t range} give also the required two-dimension surface $\Gamma_A$ in AdS$_4$ in the Poincare coordinate \eqref{eq: ads4 poin}. 

The holographic entanglement entropy is then
\begin{equation}
    S_A=\frac{area(\Gamma_A)}{4G}=\qty(\frac{1}{z_1}-\frac{1}{z_2})\frac{1}{4G}(u_a-u_b)\cot\frac{\Phi_a-\Phi_b}{2}.
\end{equation}
Compared to the entanglement entropy in CCFT$_2$ \cite{Bagchi:2014iea,Jiang:2017ecm,Apolo:2020bld,Banerjee:2024ldl}
\begin{equation}
    S_A=\frac{c_L}{6}\log(\frac{2}{\epsilon}\sin\frac{\Phi_a-\Phi_b}{2})+\frac{c_M}{12}(u_a-u_b)\cot\frac{\Phi_a-\Phi_b}{2},
\end{equation}
we have the central charges of the dual CCFT$_2$ in type III setup
\begin{equation}
    c_L=0,\ \ c_M=\frac{3}{G_N}\qty(\frac{1}{z_1}-\frac{1}{z_2}).
\end{equation}
\paragraph{Verification of the choice of $\Gamma_A$} Consider an arbitrary two-dimension surface in Poincare AdS$_4$ with the boundary $\Gamma_{a,b}$, parametrized by 
\begin{equation}
    x=f(t,z),\ \ y=g(t,z),
\end{equation}
with the boundary condition
\begin{equation}
    x_a(z)=f(t_a,z),\ \ y_a(z)=g(t_a,z),\ \  x_b(z)=f(t_b,z),\ \ y_b(z)=g(t_b,z),
\end{equation}
where $x_{a,b},y_{a,b}$ depend arbitrarily on $z$ and
$t_{a,b}$ is determined by $\phi=\phi_{a,b},u=u_{a,b}$ and the coordinate transformation \eqref{eq: flat trans}. The area functional is 
\begin{equation}
    \text{area}=\int dtdz\mathcal L,
\end{equation}
where
\begin{align}
     \mathcal L=&\frac{1}{z^2}\big(\partial_z f(t,z)^2 \partial_t g(t,z)^2-2 \partial_t f(t,z) \partial_z f(t,z) \partial_z g(t,z)
   \partial_t g(t,z)\nonumber\\
    & +\partial_t f(t,z)^2
   \partial_z g(t,z)^2-\partial_z f(t,z)^2+\partial_t f(t,z)^2-\partial_z g(t,z)^2+\partial_t g
   (t,z)^2-1\big)^{\frac{1}{2}}.
\end{align}
The Euler-Lagrange equation
\begin{equation}
   \partial_f \mathcal L =\partial_\a \frac{\partial\mathcal{L}}{\partial(\partial_a f)},\ \ \partial_g \mathcal L =\partial_\a \frac{\partial\mathcal{L}}{\partial(\partial_a g)},
\end{equation}
gives the extremal surface condition
\begin{equation}
    \partial_t^2f=0,\ \partial_z f=0,\  \partial_t^2g=0,\ \partial_z g=0.
\end{equation}
This indicates that the extremal surface is $z$ independent and it intersections with constant $z$ slices are geodesics in the flat spacetime. In other words, the extremal condition results the curves $\Gamma_{a,b}$ with $\partial_z  x_{a,b}=\partial_z y_{a,b}=0$. This reduces the problem to one dimensional lower case where the minimization procedure with respect to $x_{a,b},y_{a,b},t_{a,b}$ (or equivalently $r_{a,b}$) is the same as that in the swing surface in Flat$_3$/CCFT$_2$ correspondence \cite{Jiang:2017ecm,Apolo:2020bld}, thus gives \eqref{eq: bench equation}.

\section{Holographic correlation functions}
\label{sec:Correlation}

Next we would like to turn to holographic correlation functions in our new AdS/BCFT setups to further explore physical properties. Basically, we can generalize the computations of correlation functions based on the bulk-boundary relation \cite{Gubser:1998bc,Witten:1998qj} to our setups with flat EOW branes. We will focus on the two-point functions and one-point functions. We would like to note that green functions in AdS geometries with branes have been worked out by many earlier papers in the context of brane world holography. For this refer to e.g. \cite{Giddings:2000mu,Perez-Victoria:2001lex} for the RS2 model and \cite{Arkani-Hamed:2000ijo,Rattazzi:2000hs} for the RS1 model, where the green functions are computed and the holographic correlation functions on the branes, dual to those in the finite cut off CFT, were found. Even though our calculations below employ essentially the same green functions, we compute the holographic correlation function not on the brane but on the UV boundary, where we expect the dual CFT lives.  

\subsection{Two-point Functions in type I Case}
Bulk two-point functions in the presence of EOW branes can be computed as in the standard AdS/CFT by imposing the appropriate boundary condition at the brane. Below we consider the simple case of a bulk scalar field $\vp$ in AdS$_3$ with mass $m$ dual to an primary operator $O$ in the bulk of two dimensional BCFT. The conformal dimension of the dual operator $O$ is $\Delta=1+\nu$, where $\nu=\s{1+m^2}$. We will first calculate the two-point function in the Euclidean Poincare coordinate, where the physical region is given by $0<z<z_0$ in the coordinate $ds^2=z^{-2}(dz^2+dx_0^2+dx_1^2)$, 
and then transform it to the global coordinate.  
We simply impose the Dirichlet boundary condition at the brane $z=z_0$:
\ba
\vp(z_0)=0.  \label{DBCF}
\ea
After the Fourier transformation with respect to the coordinate $(x_0,x_1)$, the equation of motion of the scalar field can be solved generally as 
\ba
\vp(z)=z\left(\ap I_\nu(kz)+\beta I_{-\nu}(kz)\right).
\ea
By imposing the boundary condition, we obtain
\ba
\ap I_\nu(kz_0)+\beta I_{-\nu}(kz_0)=0.
\ea
Since the two-point function can be found as $\la O(k)O(-k)\lb=\frac{B}{A}$ when the bulk scalar behaves $\vp(z)\sim Az^{d-\Delta}+Bz^{\Delta}$ in AdS$_{d+1}$/CFT$_d$, we obtain
\ba
\la O(k)O(-k)\lb=-\frac{\Gamma(1-\nu)}{\Gamma(1+\nu)}\frac{I_{\nu}(kz_0)}{I_{-\nu}(kz_0)}\left(\frac{k}{2}\right)^{2\nu}.
\ea
By performing the Fourier transformation back to the real space, we find
\ba
\la O(x_0,x_1)O(x'_0,x'_1)\lb&=&\frac{1}{2\pi}\int d^2 k e^{ik(x-x')}\la O(k)O(-k)\lb\no
&=&\frac{1}{2\pi}\int^\infty_0 kdk\int^{2\pi}_0 d\theta e^{ikr\cos\theta}\la O(k)O(-k)\lb\no
&=&-\frac{\Gamma(1-\nu)}{\Gamma(1+\nu)}\int^\infty_0 kdk J_0(kr)\frac{I_{\nu}(kz_0)}{I_{-\nu}(kz_0)}\left(\frac{k}{2}\right)^{2\nu}, \label{twopws}
\ea
where we set $(x_0,x_1)=(r\cos\theta,r\sin\theta)$.

On the other hand, if we impose the Neumann boundary condition at the brane $z=z_0$:
\begin{align}
    \varphi^{\prime}(z_0)=0,
\end{align}
then we obtain
\begin{align}
    \frac{\beta}{\alpha}=-\frac{I_{-\nu}(kz_{0})+kz_{0}I^{\prime}_{-\nu}(kz_{0})}
       {I_{\nu}(kz_{0})+kz_{0}I^{\prime}_{\nu}(kz_{0})}
       =\frac{(\nu+1)I_{-\nu}(kz_0)+kz_0I_{-(\nu+1)}(kz_0)}{(\nu+1)I_{\nu}(kz_0)-kz_0I_{\nu+1}(kz_0)}.
\end{align}
This leads to
\begin{align}
    \langle O(k)\,O(-k)\rangle
&=\frac{\Gamma(1-\nu)}{\Gamma(1+\nu)}\left(\frac{k}{2}\right)^{2\nu}
\frac{I_{-\nu}(kz_{0})+kz_{0}I'_{-\nu}(kz_{0})}
     {I_{\nu}(kz_{0})+kz_{0}I'_{\nu}(kz_{0})}.
\end{align}
We can similarly obtain the real space correlation function as in (\ref{twopws}).

Now, we Wick rotate $(x_0,x_1)$ back to the Lorentzian coordinate by $x_0=it$ and $x_1=x$ and perform the conformal transformation: 
\ba
t+x=\tan\frac{\tau+\phi}{2},\ \ \ \ t-x=\tan\frac{\tau-\phi}{2},
\ea
where the cut off changes as 
\ba
e^{\rho_\infty}z_c=\frac{2}{\cos\tau+\cos\phi}.
\ea
Thus, the correlation function in the cylindrical coordinate reads
\ba
\la O(\tau,\phi)O(\tau',\phi')\lb=\frac{1}{(\cos\tau+\cos\phi)^{\Delta_+}(\cos\tau'+\cos\phi')^{\Delta_+}}\cdot \la O(t,x)O(t',x')\lb.   \label{trftw}
\ea
Note that $\la O(t,x)O(t',x')\lb$ can be found from the holographic result (\ref{twopws}) by setting
\ba
r^2=-\left(\tan\frac{\tau+\phi}{2}-\tan\frac{\tau'+\phi'}{2}\right)\left(\tan\frac{\tau-\phi}{2}-\tan\frac{\tau'-\phi'}{2}\right).
\ea
We can easily that in the short distance limit $r\to 0$, we find 
$\la O(\tau,\phi)O(\tau',\phi')\lb\propto r^{-2\Delta}$, which reproduces the standard result of CFT vacuum, as expected.

To explicitly calculate the behavior the two-point function (\ref{twopws}), we focus on the special case $m^2=-\frac{3}{4}$ which corresponds to $\nu=\frac{1}{2}$ below. Then, we get
\ba
\la O(x_0,x_1)O(x'_0,x'_1)\lb
&=&\int^\infty_0 kdk J_0(kr)\cdot\frac{k(1+e^{2kz_0})}{1-e^{2kz_0}}\no
&=&-\int^\infty_0 k^2dk \sum_{n=0}^\infty J_0(kr)\left(e^{-2nkz_0}+e^{-2(n+1)kz_0}\right)\no
&=&\sum^\infty_{n=0}\left[\frac{r^2-8n^2z_0^2}{(4n^2z_0^2+r^2)^\frac{5}{2}}+\frac{r^2-8(n+1)^2z_0^2}{(4(n+1)^2z_0^2+r^2)^\frac{5}{2}}\right],\no
&=&\sum^{\infty}_{n=-\infty}\left[\frac{3r^2}{(r^2+4n^2z_0^2)^{\frac{5}{2}}}-\frac{2}{(r^2+4n^2z_0^2)^{\frac{3}{2}}}\right].
\ea
First of all, it is straightforward to show the following behaviors
\ba
&&G(r)\simeq \frac{1}{r^3}\ \ \ (r\to 0).
\ea
By performing the Poisson re-summation, we obtain
\ba
\la O(x_0,x_1)O(x'_0,x'_1)\lb&=&\sum_{m=1}^\infty\left[
\frac{2\pi^2}{z_0^3}m^2K_2\left(\frac{\pi rm}{z_0}\right)
-\frac{4\pi m}{z_0^2r}K_1\left(\frac{\pi rm}{z_0}\right)
\right]\no
&=&\frac{2\pi^2}{z_0^3}\sum_{m=1}^\infty
m^2K_0\left(\frac{\pi rm}{z_0}\right).
\ea
This clearly shows the long distance behavior: 
\ba
\la O(x_0,x_1)O(x'_0,x'_1)\lb\sim\frac{1}{\s{r}}e^{-\frac{\pi r}{z_0}}  \label{expdec}
\ea
in the limit $r\to \infty$.

Now we turn to the behavior the two-point function in the diamond coordinate $(\tau,\phi)$. In particular, consider the limit where the location of one of the operators gets closer to the boundary  $\cos\tau+\cos\phi=0$ of the diamond. In our type I case, this boundary is acutual BCFT boundary. We note that the conformal factor (\ref{trftw}) gets divergent at the boundary.  However, $\la O(x_0,x_1)O(x'_0,x'_1)\lb$ as long as $x-x'$ is space-like, shows the exponential decay (\ref{expdec}) when $(\tau,\phi)$ approaches the boundary i.e. $r\to \infty$. Thus, the two-point function $\la O(\tau,\phi)O(\tau',\phi')\lb$ on the diamond gets vanishing in this limit. This means that the operator becomes trivial when it is moved to the null boundary and implied that the null boundary looks like a final state projection, which removes the degrees of freedom.

As a consistency check, let us compare this with the case where there is no physical boundary along the diamond, where we always have $\la O(x_0,x_1)O(x'_0,x'_1)\lb\propto r^{-2\Delta}$ and this leads 
\ba
\la O(\tau,\phi)O(\tau',\phi')\lb_{\text{vacuum}}=\frac{1}{\left(\cos(\tau-\tau')-\cos(\phi-\phi')\right)^{\Delta_+}},
\ea
which has no special behavior at the edge of diamond as opposed to our type I BCFT case and coincides with the standard result of a CFT vacuum on a cylinder.

\subsection{Two-point functions in type II Case}
Next, we address the two-point functions in the type II case. It is more convenient to perform the calculations in Poincare coordinates, where the boundary conditions are simpler. Note that the embedding \eqref{eq: embedding} is valid for both $z > 0$ and $z < 0$, with the two regions separated by the surface $X_0 = X_2$, corresponding to $z \to \pm\infty$. In the type II setup, we can place the flat EOW brane at $z = z_0 < 0$ in the negative $z$ region and examine the asymptotic behavior of the bulk scalar as $z \to 0^+$. The physical region of interest is defined by $\{0 < z < +\infty\} \cap \{-\infty < z < z_0\}$. The basic approach is as follows: first, we determine the scalar solution in the region $-\infty < z < z_0$ by imposing the boundary condition on the flat EOW brane at $z = z_0$. Then, we consider the matching condition at $z \to \pm\infty$ to determine the solution in the region $0 < z < +\infty$.

Either Dirichlet boundary condition $\vp_-(z_0)=0$ or Neumann boundary condition $\vp_-'(z_0)=0$ can be imposed. The solution in the momentum space is in general labeled as
\ba \label{eq: scalar solution -}
\vp_-(z)=z\left(\ap_- I_\nu(kz)+\beta_- I_{-\nu}(kz)\right),
\ea
where the coefficients are specified as $\ap_-,\beta_-$ since this the solution $\vp_-(z)$ valid for the region $-\infty<z<z_0$.

Then, we want to determine the matching condition at $z\to\pm\infty$. Note that the interior of the surface $X_0=X_2$ is covered by the limit \cite{Bayona:2005nq}
\begin{equation}\label{eq: poin limit}
    z\to\pm\infty,\ \ z^2+r^2=\frac{a}{2}z,\ \ x^2=b^2z^2,\ \ r^2=x^2-t^2,
\end{equation}
in Poincare coordinate, related to the global coordinate by
\begin{equation}
    \sec\rho=\sqrt{1+a^2+b^2},\ \ \cos\tau=\frac{a}{\sqrt{1+a^2+b^2}},\ \  \sin\phi=\frac{a}{\sqrt{a^2+b^2}}.
\end{equation}
Then, we perform the Fourier transformation on the solution to go back to the real space
\begin{equation}\label{eq: scalar solution real}
    \vp_{-,+}(z,r)=\frac{1}{2\pi}\int^\infty_0 kdk\int^{2\pi}_0 d\theta e^{ikr\cos\theta}\vp_{-,+}(z)=\int^\infty_0 kdkJ_0(kr)\vp_{-,+}(z),
\end{equation}
where $\vp_+$ labels the solution in the region $0<z<+\infty$
\ba \label{eq: scalar solution +}
\vp_+(z)=z\left(\ap_+ I_\nu(kz)+\beta_+ I_{-\nu}(kz)\right),
\ea
where the coefficients are labeled as $\ap_+,\beta_+$. We impose the matching condition formally
\begin{equation}
    \lim_{z\to+\infty}\vp_+(z,r)|_{z^2+r^2=\frac{a}{2}z}=\lim_{z\to-\infty}\vp_-(z,r)|_{z^2+r^2=\frac{a}{2}z}.
\end{equation}
In the limit \eqref{eq: poin limit},
\begin{equation}
    \vp_{-,+}\to\int _{0}^\infty kdk I_0(-kz+\frac{ak}{4}) z\qty(\ap_{-,+} I_\nu(kz)+\beta_{-,+} I_{-\nu}(kz)).
\end{equation}
With the large $z$ behavior of $I_\nu(z)$ for fixed $\nu$,
\begin{equation}
I_\nu(z)\sim\begin{cases}
   { e^z\over \sqrt{2\pi z} },&\ \ |arg\  z|<\frac{1}{2}\pi-\delta,\\
   { e^{-z+(\nu+\frac{1}{2})\pi i}\over \sqrt{2\pi z}},&\ \ \frac{1}{2}\pi+\delta<\pm arg\  z<\frac{3}{2}\pi-\delta,\\
\end{cases}
\end{equation}
Since the solutions \eqref{eq: scalar solution real} get divergent exponentially as $z\to \pm\infty$, we should expand the solutions in terms of $z$ at $z\to\pm\infty$ and match the coefficients of the divergent term and the constant term, leading to 
\begin{equation}
 \ap_+=-\ap_- e^{8 a k-i \pi  v},\ \ \beta_+=-\beta_-e^{8 a k+i \pi  v}.
\end{equation}
Note that the ratio $\frac{\beta_{-,+}}{\ap_{-,+}}$ is $k$-independent, so that we find the relation between the two-point function $\langle O(t,x)O(t',x')\rangle_{II}$ in the type II case and that $\langle O(t,x)O(t',x')\rangle_{I}$ in the type I case
\begin{equation}
    \langle O(t,x)O(t',x')\rangle_{II}=e^{2\pi i\nu}\langle O(t,x)O(t',x')\rangle_{I}.
\end{equation}
The difference is only the overall phase factor $e^{2\pi i \nu}$. We can then go to the cylindrical coordinate where this overall phase factor cancels the signs in the conformal factor $(\cos\tau+\cos\phi)^{\Delta_+}(\cos\tau'+\cos\phi')^{\Delta_+}$ and get
\begin{equation}
    \langle O(\tau,\phi)O(\tau',\phi')\rangle_{II}= \langle O(\tau,\phi)O(\tau',\phi')\rangle_{I}.
\end{equation}

\subsection{One-point functions in type I and II case}
We will calculate the one-point functions in the type I and II setups in this subsection. To have a non-vanishing one-point function in the AdS/BCFT as considered in \cite{Fujita:2011fp,Suzuki:2022xwv}, the source term is added on the EOW brane
\begin{equation}
    I_{s}=-\frac{a}{8\pi G}\int_{EOW}\sqrt{|h|}\phi,
\end{equation}
where $h$ is the determinant of the induced metric on the EOW brane and $a$ is the coupling constant. In this case, the holographic one-point function on the asymptotic boundary can be derived using the bulk-to-boundary propagator as follows \cite{Kastikainen:2021ybu,Suzuki:2022xwv,Izumi:2022opi}:
\begin{equation}\label{eq:Onepoint}
  \langle{O(x)}\rangle = a \cdot \, \int_{\mt{EOW}} d^2 x_b \sqrt{|h|} K_{\mt{Bb}}^\Delta (x ; x_b) \,.
\end{equation}
where the coordinates $x$ and $x_b$ denote the two points on the asymptotic boundary and the EOW brane, respectively and $K_{\mt{Bb}}^\Delta (x ; x_b)$ is the bulk-to-boundary propagator with the EOW brane for the scalar operator with conformal dimension $\Delta$. It can be related to the bulk-to-bulk propagator in AdS$_3$ without the EOW  brane\cite{Witten:1998qj,Freedman:1998tz,Ammon:2015wua} 
\begin{equation}
G_{bulk-bulk}(x,x')=\frac{\Gamma(\Delta)}{2^{\Delta+1}( \Delta-1)\pi\Gamma\left(\Delta-1\right)} \xi^{\Delta} \cdot{ }_2 F_1\left(\frac{\Delta}{2}, \frac{\Delta+1}{2} ; \Delta- ; \xi^2\right) \,,
\end{equation}
where we have introduced the chordal distance $\xi$ that is associated with the geodesic distance $D(x,x')$ 
\begin{equation}
D(x ; x') =  \, \ln \left(\frac{1+\sqrt{1-\xi^2}}{\xi}\right) \,, \qquad  \xi = \frac{1}{\cosh  \( D(x ; x')  \)}.
\end{equation}
The bulk-to-bulk propagator satisfies the equation
\begin{equation}\label{eq:EOMGBB}
\left(\square_G +m^2\right) G_{\mt{BB}}^\Delta(x ; x')=\frac{\delta^{(d+1)}(x- x')}{\sqrt{G}}\,,
\end{equation}
where $\square_G$ is the scalar Laplacian operator in AdS spacetime and the mass of the bulk field is related to the conformal dimension of the dual operator on the boundary via the standard AdS/CFT dictionary $m^2 = \Delta (\Delta -2)$.

In the presence of a flat EOW brane, the bulk-to-bulk propagator can be calculated using the method of images, with the brane acting as a mirror. The Dirichlet or Neumann boundary conditions on this brane can be replicated by introducing an additional source in the equation \eqref{eq:EOMGBB}, positioned according to the reflection of the brane. Additionally, the background AdS spacetime must also be reflected by this brane. Thus, the bulk-to-bulk propagator with the flat EOW brane is
\begin{equation}
    G_{\text{EOW}}(x,x')=G_{BB}(x,x')\pm \tilde G_{BB}(x,Rx')
\end{equation}
where $\tilde G_{BB}(x,Rx')$ is calculated in the glued AdS geometry via the EOW brane and $Rx'$ is the location of the image of $x'$. The sign $\pm$ is suitable for Neumann and Dirichlet boundary condition. Specially, if one of the bulk point is on the EOW brane,
\begin{equation}
    \tilde G_{BB}(x,Rx')|_{x'\in\text{EOW}}=G_{BB}(x,x')|_{x'\in\text{EOW}}.
\end{equation}
So only the Neumann boundary condition leads to non-vanishing bulk-to-bulk propagator with the flat EOW brane, if one of the bulk point is on the brane,
\begin{equation}\label{eq: bulk-to-bulk eow}
    G_{\text{EOW}}(x,x')|_{x'\in\text{EOW}}=2G_{BB}(x,x')|_{x'\in\text{EOW}}
\end{equation}
Then, the bulk-to-boundary propagator with the EOW brane appearing in \eqref{eq:Onepoint} $K_{\mt{Bb}}^\Delta (x ; x_b)$ can be derived by taking the boundary limit of the bulk-to-bulk propagator \eqref{eq: bulk-to-bulk eow} with the flat EOW brane. For example, in the Poincare coordinate, we have 
\begin{equation}
   K_{\mt{Bb}}^\Delta (x ; x_b)=2\lim_{z\to0}\frac{2\Delta-2}{z^\Delta}G_{\text{EOW}}(x,x_b)|_{x_b\in\text{EOW}}.
\end{equation}
The bulk-to-boundary propagator in other coordinates is obtained from the coordinate transformations.

We then consider the large $\Delta$ limit of these operators, since the disconnected phase contribution to the holographic entanglement entropy is related to the one-point function of the twist operator with large $\Delta$. By taking $\Delta \to \infty$, we can approximate the bulk-to-bulk propagator and by the geodesic distance \cite{Balasubramanian:1999zv},
\begin{equation}
    G_{BB}(x,x')=\frac{1}{2\Delta-2}e^{-\Delta D(x ; x')}\qty(\frac{\Delta}{2\pi}\frac{1+\sqrt{1-\xi^2}}{\sqrt{1-\xi^2}}),
\end{equation}
resulting the bulk-to-boundary propagator with the flat EOW brane in the Poincare coordinate,
\begin{equation}\label{eq: bulk-to-boundary poincare}
     K_{\mt{Bb}}^\Delta (x ; x_b)=2\frac{\Delta}{\pi}\qty(\frac{z_0}{z_0^2+(x-x_b)^2-(t-t_b)^2})^\Delta.
\end{equation}
\paragraph{Type I setup} In this case, the brane is located at $z=|z_0|>0$. In the Poincare patch, equations \eqref{eq:Onepoint},\eqref{eq: bulk-to-boundary poincare} can be used directly to calculate the holographic one-point function with large $\Delta$,
\begin{equation}\label{eq: 1pt poin I}
    \langle O(x)\rangle_{I}=2a\int dx_bdt_b \frac{\Delta}{\pi z_0^2}\qty(\frac{z_0}{z_0^2+(x-x')^2-(t-t_b)^2})^\Delta=\frac{a\Delta}{|z_0|^\Delta}.
\end{equation}
In the global coordinate, it is more connivent to use the geodesic length with one global coordinate and one Poincare coordinate, e.g.\eqref{eq: geo glo poin}, in the geodesic length approximation,
\begin{equation}
     K_{\mt{Bb}}^\Delta (x ; x_b)=2 e^{-\Delta D_{13}(x ; x_b)}\qty(\frac{\Delta}{\pi}),
\end{equation}
leading to
\begin{equation}
    \langle O(x)\rangle_{I}=2a\int dx_bdt_b \frac{1}{z_0^2}K_{\mt{Bb}}^\Delta (x ; x_b)=a\Delta\qty(\frac{2}{|z_0|(\cos\tau+\cos\phi)})^\Delta.
\end{equation}
This is consistent with the result via the plane-to-cylinder conformal map and the Poincare coordinate calculation \eqref{eq: 1pt poin I}, serving as the consistency check of the validity of the chordal distance method to get the geodesic with one global coordinate and one Poincare coordinate.

\paragraph{Type II setup} In this case, the flat EOW brane is located at $z=-|z_0|<0$. For large $\Delta$, the geodesic length \eqref{eq: geo glo poin} can be exploited to get
\begin{equation}\label{eq: 1pt II}
    \langle O(x)\rangle_{II}=2a\int dx_bdt_b \frac{1}{z_0^2}K_{\mt{Bb}}^\Delta (x ; x_b)=a\Delta\qty(\frac{2}{-|z_0|(\cos\tau+\cos\phi)})^\Delta.
\end{equation}
Compared to the type I setup with the brane at $z=|z_0|$, there is an overall phase factor in the one-point function $e^{\pi i\Delta}$, which is consistent with the phase factor in front of the two-point function $e^{2\pi i\nu}=e^{2\pi i\Delta}$.

\paragraph{Saddle point approximation and disconnected phase contribution to holographic entanglement entropy} Note that the integration \eqref{eq: 1pt poin I}, \eqref{eq: 1pt II} above to get the one-point function can also be calculated by the saddle point approximation, with the saddle point
\begin{equation}
    x_b=x,\ \ t_b=t,
\end{equation}
in the Poincare coordinate calculation and
\begin{equation}
    t_b=\frac{ \sin \tau}{\cos \tau+\cos \phi},\ \ \ x_b=\frac{ \sin \phi}{\cos \tau+\cos \phi},
\end{equation}
in the one Poincare coordinate and one global coordinate calculation. They are actually the same saddle point as \eqref{eq: sol dis I}, \eqref{eq: II saddle} in the disconnected phase contribution to the holographic entanglement entropy. From the BCFT point of view, the entanglement entropy can be derived from the two-point function of the twist operator,
\begin{equation}\label{EErep}
S_{\mathcal{A}} \equiv \lim_{n\to 1}\frac{1}{1-n}\log \la \sigma_n(A_1)\bar{\sigma}_n(A_2)\lb \,,  
\end{equation} 
where the normalization of twist operators is fixed as $\la \sigma_1(A_1)\bar{\sigma}_1(A_2)\lb=1$ and their conformal dimension is $\Delta_n=\frac{c}{12}\qty(n-\frac{1}{n})$. In the semi-classical limit of a holographic field theory, this is equivalent to taking a large central charge limit
$\Delta_n \sim c  \sim \frac{1}{\GN} \to \infty$. For a holographic BCFT, the two-point functions at leading order in the large central charge limit $c\to\infty$ are dominated by two distinct channels 
\begin{equation}\label{eq:twistors}
\ev{\sigma_n(A_1)\bar{\sigma}_n(A_2)}_{\mathrm{BCFT}}= \max 
  \begin{cases} \ev{\sigma_n(A_1)\bar{\sigma}_n(A_2)}  \\
\ev{\sigma_n(A_1)}\ev{\bar{\sigma}_n(A_2)}\\
  \end{cases} \,.
 \end{equation}
Here, the one-point function is non-vanishing due to the presence of a boundary in the background. These two channels correspond to the connected and disconnected phase contributions to the holographic entanglement entropy. In the disconnected phase, the one-point function of the twist operator with large $\Delta$ can be calculated by the saddle point approximation above
\begin{equation}
    \langle{\sigma_n(A)} \rangle \sim \sqrt{\frac{|h|}{ \det (D''(x_A ; x_b^\ast))}}\; e^{-\Delta_n{D}(x_A; x_b^\ast)} \,= e^{-\Delta_n{D}(x_A; x_b^\ast)} \,,
\end{equation}
leading to the contribution to the holographic entanglement entropy from the disconnected geodesics
\begin{equation}\label{formulaHEER}
\begin{split}
S_{\mathcal{A}} &\approx  \lim_{n\to 1}\frac{1}{1-n}\( \lim_{\Delta_n \to \infty} \( \log \ev{\sigma_n(A_1)} + \log \ev{\sigma_n(A_2)}  \) \)     \\ 
&=  \frac{c}{6} \, \(    D_{13}  +D_{24}  \)  \,, \\
\end{split}
\end{equation}
where the point $A_3,A_4$ on the flat EOW brane is determined by the saddle point. Note that this is similar to the usual AdS$_3$/BCFT$_2$ correspondence with an AdS brane, where the equivalence between \eqref{eq:twistors} and the holographic entanglement entropy formula can be demonstrated using the geodesic approximation and different from the dS EOW brane case \cite{Hao:2024nhd}, where the saddle point is not stable.

\subsection{Type III Case}
Finally, we turn to the type III case.
In the Poincare coordinate, the gravity dual is situated in the region $z_1<z<z_2$ surrounded by the two EOW branes. Consider a free massive scalar $\vp$ in the type III geometry.  In the metric (\ref{poinc}), the equation of motion reads
\begin{align}
    z^{d+1} \partial_z\left(z^{1-d} \partial_z \varphi\right)+z^2\left(-\partial_t^2+\partial_{\vec{x}}^2\right) \varphi-m^2 R^2 \varphi=0.
\end{align}
The Fourier transformation
\begin{align}
\vp(t,x,z) = \psi(z)\,e^{-i\omega t + i \vec{k}\cdot \vec{x}},   
\qquad
q^2 \equiv \omega^2 - k^2, \label{wave}
\end{align}
leads to the radial ODE
\begin{align}
    z^2 \psi^{\prime \prime}(z)-(d-1) z \psi^{\prime}(z)+\left(q^2 z^2-m^2 R^2\right) \psi(z)=0.
\end{align}
The above is the Bessel equation of order
$\nu=\sqrt{\qty(\frac{d}{2})^2+m^2R^2}$. 
Its general solution takes the form:
\begin{align}
\psi(z)=z^{d / 2}\left[A J_\nu(q z)+B Y_\nu(q z)\right] .
\end{align}
We impose Neumann conditions on the scalar at $z=z_1$ and $z=z_2$:
\begin{align}
\psi'(z_{1})=0,\qquad \psi'(z_{2})=0.
\end{align}
Define the shorthand
\begin{align}
  F_d(x) \equiv \frac{d}{2} J_\nu(x)+x J_\nu^{\prime}(x), \quad G_d(x) \equiv \frac{d}{2} Y_\nu(x)+x Y_\nu^{\prime}(x).
\end{align}
Then, the general solution can be obtained:
\begin{align}
  \psi^{\prime}(z)=z^{\frac{d-2}{2}}\left[A F_d(q z)+B G_d(q z)\right].
\end{align}
The Neumann conditions at $z=z_{1}$ and $z=z_{2}$ give a homogeneous linear system
\begin{align}
\begin{pmatrix}
F_d(qz_{1}) & G_d(qz_{1})\\
F_d(qz_2) & G_d(qz_2)
\end{pmatrix}
\begin{pmatrix}A\\B\end{pmatrix}
=0,
\end{align}
which admits a nontrivial solution only if its determinant vanishes:
\begin{align}
F_d(qz_{1})\,G_d(qz_{2})
\;-\;
G_d(qz_{1})\,F_d(qz_{2})
=0.
\end{align}
This is the quantization condition determining the discrete spectrum $q=q_{n}$.\\
A convenient choice of coefficients that automatically satisfies $\psi'(z_{1})=0$ is
\begin{align}
A=G_d\bigl(q_{n}z_{1}\bigr),
\quad
B=-\,F_d\bigl(q_{n}z_{1}\bigr).
\end{align}
Thus, the properly modes are
\begin{align}
\psi_{n}(z)
=
z^{\frac{d-2}{2}}\Bigl[
J_{\nu}(q_{n}z)\,G_d(q_{n}z_{1})
\;-\;
Y_{\nu}(q_{n}z)\,F_d(q_{n}z_{1})
\Bigr],
\end{align}
and one checks directly that $\psi'_{n}(z_{1})=\psi'_{n}(z_{2})=0$.  \\
This calculation shows that we can view the scalar field in the $d+1$ dimensional bulk as the tower of infinitely many scalar field with various masses $q_n^2$ in $d$ dimensional flat space R$^{1,d-1}$, as we can find from (\ref{wave}). Thus, the analysis of the correlations functions will essentially become the same as that in the flat spacetime holography \cite{Hijano:2017eii,Hijano:2019qmi,Donnay:2022wvx,Nguyen:2023miw}. We leave its detailed analysis for a future work.
Instead, below we will see how the spectrum looks like in a few examples.

\paragraph{$d=3$ (AdS$_4$) case}
We set $d=3$ and $m^2=0$, namely $\nu=\frac32$. The Neumann boundary conditions at $z=z_1,z_2$ are built from
\begin{align}
    F_3(x) &= \frac{3}{2} J_\frac32(x)+x J_\frac32^{\prime}(x) ,~~~
    \quad G_3(x) = \frac{3}{2} Y_\frac{3}{2}(x)+x Y_\frac{3}{2}^{\prime}(x).
\end{align}
Using the Bessel identity $xJ'_\nu+\nu J_\nu=xJ_{\nu-1}$ (and the same for $Y_\nu$), one immediately finds
\begin{align}
    F_3(x)=xJ_{\frac12}(x),\qquad
    G_3(x)=xY_{\frac12}(x).
\end{align}
Therefore, the quantization condition can be written as
\begin{align}
    J_{\frac12}\left(q z_1\right) Y_{\frac12}\left(q z_2\right)-Y_{\frac12}\left(q z_1\right) J_{\frac12}\left(q z_2\right)=0 .
\end{align}
For half–integer order, the Bessel functions admit closed forms
\begin{align}
    J_{\frac12}(x)=\sqrt{\frac{2}{\pi x}}\sin{x},~~Y_{\frac12}(x)=-\sqrt{\frac{2}{\pi x}}\cos{x}
\end{align}
and all common prefactors cancel out in the determinant.
\begin{align}
    -\sin \left(q z_1\right) \cos \left(q z_2\right)+\cos \left(q z_1\right) \sin \left(q z_2\right)=\sin \left(q\qty(z_2-z_1)\right)=0 .
\end{align}
Writing $L\equiv z_2-z_1>0$, the spectrum is thus
\begin{align}
    q_n=\frac{n\pi}{L}~~~(n\in\mathbb{Z}).
\end{align}
\paragraph{$d=2$ (AdS$_3$) case}
We set $d=2$ and $m^2=-\tfrac{3}{4}$, namely $\nu=\frac12$. Proceeding as above and
using the half–integer closed forms for $J_{1/2},Y_{1/2}$, the
 quantization condition reduces exactly to
\begin{align}
    \left(q^2 z_1 z_2+\frac{1}{4}\right) \sin (q L)+\frac{1}{2} q L \cos (q L)=0
\end{align}
Equivalently,
\begin{align}
    \tan qL=-\frac{\frac12 qL}{q^2z_1z_2+\frac14}
\end{align}
The solution of this equation for $n\gg1 $ is given as follows:
\begin{align}
    q_n=\frac{n \pi}{L}-\frac{L}{2 z_1 z_2 n \pi}+O\left(\frac{1}{n^3}\right).
\end{align}

\section{On-shell action in Euclidean Poincare AdS$_3$}
\label{sec:Action}
In this section, we will calculate the Euclidean on-shell action of Einstein gravity in AdS$_3$ ($d=2$), which is expected to be equal to the partition function (or free energy) of the dual field theory. 
The starting point is the Euclidean Poincare AdS$_3$,
\begin{equation}
    ds^2=\frac{dt_E^2+dx^2+dz^2}{z^2},
\end{equation}
with the EOW brane
\begin{equation}
    z=z_0.
\end{equation}
The Euclidean on-shell action is
\begin{equation}\label{eq: gravity action}
    I_E=-\frac{1}{16\pi G_N}\int_{M} \sqrt{g}({\cal R}-2\Lambda)-\frac{1}{8\pi G_N}\int_{Q}\sqrt{h}(K-T),
\end{equation}
where $G_N$ is the three-dimensional gravitational constant. 
The first term represents the Einstein-Hilbert term calculated in the bulk, specifically within the region \eqref{eq: types} defined in AdS$_3$. The second term corresponds to the Gibbons-Hawking terms evaluated on the boundary, which includes the surfaces at $z=0$ and $z=z_0$ in the type I case, and the brane at $z=z_0$ in the type II case. Besides, the brane tension term is considered to give the proper EOW brane equation of motion \eqref{eq: brane eom}. Below we will mainly focus on the type I case as a similar result can also be obtained for type II by flipping the sign. Below we will give two different prescriptions to regulate the UV and IR divergences, which eventually gives the same on-shell action. 

\subsection{Regularization by deforming the cut off surface}
Direct calculations encounter divergences from the $z=z_0$ surface, representing the UV divergence typical in calculations involving the asymptotic boundary of AdS, as well as an additional IR divergence as $t_E, x$ approach infinity. To address these issues, it is crucial to introduce a cutoff surface to regulate both types of divergences. Subsequently, we calculate the action \eqref{eq: gravity action} within the region bounded by the EOW brane at $z=z_0$ and the cutoff surface.

We deform the cut off surface into the form
\begin{equation}\label{eq: cutoff IRUV}
    r^2+\qty(z-\frac{1}{\epsilon}-\epsilon)^2=\frac{1}{\epsilon^2},
\end{equation}
to regulate the IR divergence, where
\begin{equation}
    r^2=x^2+t_E^2.
\end{equation}
and $\epsilon$ is a small parameter. This surface is a very large semi-sphere with radius $\frac{1}{\epsilon}$, whose center is located at 
\begin{equation}
    z=\e+\frac{1}{\e},\ \ x=t_E=0.
\end{equation}
It intersects with $z=const.$ slices at circles with $r^2=-\frac{(\e-z) \left(\e^2-\e z+2\right)}{\e}\to \frac{2z}{\e}$ so that it regulate the IR divergence. As $\e\to0$, the cutoff surface becomes the usual cutoff surface
\begin{equation}
    z=\e,
\end{equation}
so that it regulate the UV divergence from $z=0$ at the same time. Then, we calculate the on-shell action in the region bounded by the EOW brane and the cutoff surface,
\begin{equation}
    I_E=\frac{\left(\epsilon ^2+1\right) (z_0-\epsilon )+z_0 \epsilon ^2 (\log \epsilon -\log z_0)}{4 G z_0 \epsilon ^2},
\end{equation}
where the contribution from the Gibbons-Hawking term on the cutoff surface is also included whose trace of the extrinsic curvature is $ K_\epsilon=2(1+\epsilon^2)$.

We can further subtract the leading $\frac{1}{\epsilon^2}$ and subleading $\frac{1}{\epsilon}$ divergence by further adding the counter term in the gravitational action \eqref{eq: gravity action},
\begin{equation}
    I_{ct}=\frac{1}{8\pi G_N}\int_{\epsilon}\sqrt{h_{\epsilon}}\qty(1-\frac{{\cal R}_\epsilon}{4}\log\epsilon).
\end{equation}
This is the usual counter term to cancel the divergence, calculated on the cutoff surface where $h_\epsilon$ is the induced metric on the cutoff surface and ${\cal R}_\epsilon$ is Ricci scalar of this surface which vanishes as $\epsilon\to0$. For the surface \eqref{eq: cutoff IRUV}, the counter term reads
\begin{equation}
    I_{ct}=-\frac{\left(\epsilon ^2+1\right) (z_0-\epsilon )}{4 G_N z_0 \epsilon ^2},
\end{equation}
leading to the total on-shell action in the type I:
\begin{equation}\label{eq: tot action}
    I_{tot,I}=I_E+I_{ct}=-\frac{1}{4G_N}\log\frac{z_0}{\epsilon}.
\end{equation}
What we left is the $\log\epsilon$ divergence only, similar to the usual calculation in pure AdS$_3$. 

We also find the total on-shell action in the type II case as
\begin{equation}\label{eq: tot action II}
    I_{tot}=-\frac{1}{4G_N}\log\frac{z_0}{\epsilon}.
\end{equation}
This is found just by flipping the sign of the type I result because the bulk region of type II is the complement of that of type I.

\subsection{Regularization by deforming flat EOW branes to AdS ones}
We can also consider another regularization to get the on-shell action in type I configuration. Note that the flat EOW brane can be regarded as the limit from the AdS EOW brane,
\begin{equation}\label{eq: ads brane}
    t_E^2+x^2+(z+r_0-z_0)^2=r_0^2,\ \ r_0>\frac{z_0}{2},
\end{equation}
with the Ricci scalar 
\begin{equation}
    R=\frac{2 z_0 (z_0-2 r_0)}{r_0^2},
\end{equation}
and the trace of the extrinsic curvature
\begin{equation}
    K=2T=-2+\frac{2z_0}{r_0}.
\end{equation}
As the limit $r_0\to\infty$, it recovers the flat EOW brane $z=z_0$. In this case, the cutoff surface is imposed at $z=\epsilon$ as usual. A parallel
calculation is carried out resulting
\begin{equation}
    I_E=\frac{1}{4G_N}\qty(\frac{2 r_0 z_0-z_0^2}{2 \epsilon
   ^2}+\frac{z_0-r_0}{\epsilon }-\log
   \left(\frac{z_0}{\epsilon }\right)-\frac{1}{2}).
\end{equation}
Firstly, the divergence as $z\to 0$ is regulated by setting $z=\epsilon$. Secondly, the divergence due to the $r_0 \to \infty$ limit for a fixed $z_0$, which corresponds to the IR divergence, is also present. Furthermore, the bulk divergence is $\frac{1}{\epsilon^2}$, while the boundary term contributes $\frac{1}{\epsilon}$.

We should subtract the divergence with the additional counter term in the action calculated at $z=\epsilon$
\begin{equation}
    I_{ct}=\frac{1}{4G_N}\qty(-\frac{r_b^2}{2\epsilon^2}-\frac{z_0^2-r_b^2}{2z_0\epsilon}+\frac{1}{2}).
\end{equation}
where $r_b$ is the radius of circle 
\begin{equation}
    r_b^2\equiv\tau^2+x^2=2r_0 z_0-z_0^2
\end{equation}
which is the intersection between the AdS brane \eqref{eq: ads brane} and the asymptotic boundary $z=0$. Then 
\begin{equation}
    I_{tot,I}=-\frac{1}{4G_N}\log\frac{z_0}{\epsilon}.
\end{equation}
This gives the same result as \eqref{eq: tot action} from the flat EOW brane directly and the cutoff surface \eqref{eq: cutoff IRUV}, serving as a consistency check.

\subsection{Comparison with boundary entropy}

For AdS EOW branes \cite{Takayanagi:2011zk,Fujita:2011fp}, the boundary entropy (or $g$-function) is related to its tension via:
\begin{equation}
    \log g=\frac{c}{6}\log\sqrt{\frac{1+T}{1-T}}.
\label{loggg}
\end{equation}
For the tensionless AdS EOW brane $T=0$, we have $\log g=0$, which corresponds to $r_b=z_0$ (or equally $r_0=z_0$) in (\ref{eq: ads brane}). On the other hand, in the type I (and type II) flat EOW brane limit $T=-1$ (and $T=1$) we have $\log g=-\infty$ (and  $\log g=\infty$). Since this $g$ function is defined as the regularized disk partition function, $-\log g$ should coincide with the regularized on-shell action in the gravity. 

In our calculation based the AdS brane (\ref{eq: ads brane}), 
we can regularize the UV divergence as  
\ba
 I_{ren,I}=I_{tot,I}-I_{disk(0)}=-\frac{1}{4G_N}\log\frac{z_0}{r_b},
\ea
where the $I_{disk(0)}$ is the contribution from the un-regularized disk partition function for $T=0$ at the radius $r_b$, given by $I_{disk(0)}=-\frac{1}{4G_N}\log \frac{r_b}{\ep}$. 

In the actual type I setup, we need to take the limit $r_b\to\infty$ and this leads to $I_{ren,I}=\infty$. Similar we obtain $I_{ren,II}=-\infty$. Indeed, they agree with the expected behaviors of $\log g$ (\ref{loggg}) in $T=-1$ and $T=-1$ limit, respectively.


\section{Euclidean Holography}
\label{sec:Euclid}

Finally let us turn to a possible holographic interpretation of the Euclidean setup of the AdS/BCFT with a flat EOW brane. As we noted in section \ref{sec:setups}, the EOW brane (\ref{EEOW}) in the Euclidean global AdS$_{d+1}$ intersects with the AdS boundary $\rho\to\infty$ only at a point $\phi=\pi$ and $\tau_E=0$. We will focus on the type II setup 
as illustrated in Fig.\ref{fig:EAdS}.

\subsection{Holographic Setup}
\begin{figure}[ttt]
		\centering
		\includegraphics[width=10cm]{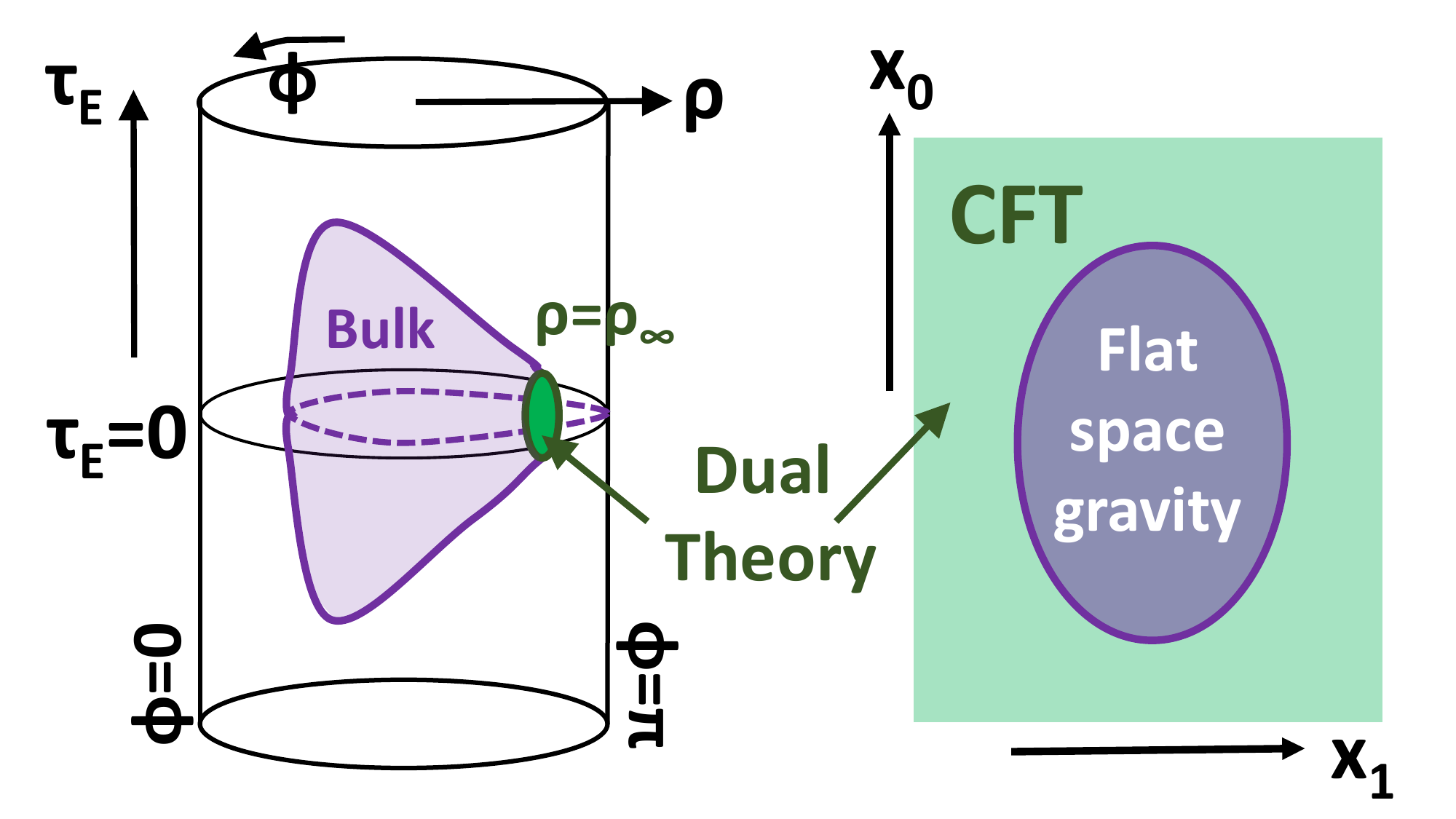}
		\caption{A sketch of Euclid type II set up with a UV cut off (left) and its dual field theory description via the double holography (right) in the Poincare coordinate.} 
		\label{fig:EAdS}
\end{figure}

In this setup, the $d$ dimensional gravity on a Euclidean flat space on the brane $z=z_0$ together with the $d+1$ bulk gravity in the region surrounded by the brane, is expected to be equivalent to a certain large $N$ theory localized on the point $\phi=\pi$ and $\tau_E=0$. If we introduce the UV cut off $\rho\leq \rho_\infty$, then the dual large $N$ theory lives on a tiny region given by a $d$ dimensional round ball, given by
\ba
\tau_E^2+(\pi-\phi)^2\leq \frac{4R}{z_0}e^{-\rho_\infty},
\label{dualcir}
\ea
on the cut off boundary $\rho=\rho_\infty$ of the global AdS$_{d+1}$. Refer to the left panel of Fig.\ref{fig:EAdS} for a sketch. 

If we focus on the $d$ dimensional flat space gravity realized on the brane, this is dual to the $d-1$ dimensional boundary of the ball region (\ref{dualcir}). In this way, by embedding the flat brane inside AdS, we can find an explicit example of holography for gravity on a flat Euclidean space. 

In terms of the Poincare AdS (\ref{poinc}), the 
flat EOW brane has the UV cut off as
\ba
(R^2+z^2+x^2+x_0^2)^2-4R^2x_0^2\le 4R^2z^2\cosh^2 \rho_\infty, \label{regionw}
\ea
with $z$ set as $z=z_0$, where $x_0=it$ is the Euclidean time coordinate.
As in the right panel of Fig.\ref{fig:EAdS}, 
there is a flat space gravity in the above region (\ref{regionw})
in the dual field theory via the double holography, coupled to the CFT situated outside.

\subsection{On-shell action}
Since there is no space and time in the dual theory, we would like to compute its partition function by evaluating the gravity on-shell action below as the only physical quantity.
Notice that this calculation is different from the one in the previous section. Here we now impose the UV cut off $\rho\le \rho_\infty$ in the global AdS to examine the leading UV divergent contributions, while in the previous section we wanted to extract the boundary entropy contribution in the Poincare patch.

The gravity action is given by (\ref{eq: gravity action}).
In the current type II setup, we have
\ba
&& {\cal R}=-\frac{d(d+1)}{R^2}, \ \ \Lambda=-\frac{d(d-1)}{2R^2},\no
&& K=\frac{d}{R},\ \ \ T=\frac{d-1}{R}.
\ea
Thus, we can write it as 
\ba
 I_E=\frac{d}{8\pi G_N R^2}\int_{M_{d+1}} \s{g}
 -\frac{1}{8\pi G_N R}\int_{Q_d} \s{h}.
\ea
To evaluate this in the Poincare coordinate, let us define $V(z)$ to be the volume of the regions $(x_0,x_1,\ddd,x_{d-1})$ which satisfy (\ref{regionw}) for a fixed value of $z$. In the presence of the UV cut off $\rho\leq \rho_\infty$, the coordinate $z$ in the gravity dual takes the values $z_0\leq z\leq Re^{\rho_\infty}$. Thus, $I_E$ is expressed as 
\ba
I_E=\frac{d}{8\pi G_N R}\int^{R e^{\rho_\infty}}_{z_0} dz\frac{V(z)}{z}
-\frac{V(z_0)}{8\pi G_N R}.
\ea
The function $V(z)$ is evaluated as 
\ba
V(z)=V_{d-2}\int^{x^{max}_0}_0 dx_0
\int^{r^{max}(x_0)}_0 r^{d-2}dr,
\ea
where $V_{d-2}$ is the volume of the unit $d-2$ dimensional sphere and we introduced 
\ba
&& x^{max}_0=\s{R^2+2Rz\sinh\rho_\infty-z^2},\no
&& r^{max}(x_0)=\s{-(x_0^2+R^2+z^2)+2R\s{x_0^2+z^2\cosh\rho^2_\infty}}.
\ea
In the UV limit $\rho_\infty\to \infty$, we can  approximate $x^{\max}$ and $r^{\max}$ so that they are on the $d$ dimensional round disk:
\ba
(x^{\max})^2+(r^{max})^2\leq  Rze^{\rho_\infty},
\ea
This lead to the following estimation:
\ba
V(z)\simeq \frac{V_{d-1}}{d}R^{\frac{3}{2}d}z^{-\frac{d}{2}}e^{\frac{d}{2}\rho_\infty}.
\ea
Finally, we can evaluate the on-shell action as 
\ba
I_E=\frac{R^{d-1}}{8\pi d G_N}\left(\frac{R}{z_0}\right)^{\frac{d}{2}}e^{\frac{d}{2}\rho_\infty}+\ddd,  \label{freenergy}
\ea
where the omitted terms are in the higher order with respect to the UV cut off.

In the standard holographic interpretation of the global AdS, the UV cut off $\ep$ of the dual CFT is given by $\ep\sim e^{-\rho_\infty}$. In this interpretation, the partition function $Z=e^{-I_E}$ with (\ref{freenergy}) looks like that in a $\frac{d}{2}$ dimension CFT. This apparent fractional dimension arises from characteristic profile of the flat EOW brane. As $z_0$ gets smaller,  the region of the flat space where the gravity is present gets expanding, which is give by a $d$ dimensional disk whose the radius is given by $\s{Rz_0}e^{\frac{\rho_\infty}{2}}$. This is consistent with the fact that  (\ref{freenergy}) is proportional to $\left(\frac{R}{z_0}\right)^{\frac{d}{2}}$, which may be viewed as a measure of the degrees of freedom. 

Our analysis here implies that the gravity dual of $d$ dimensional Euclidean flat Ball $B^d$ with the radius $\s{Rz_0}e^{\frac{\rho_\infty}{2}}$ is given by a certain field theory on another $d$ dimensional ball with the radius $2\s{\frac{R}{z_0}}e^{\frac{\rho_\infty}{2}}$ as follows from (\ref{dualcir}).
Indeed if we assume the UV cut off $\ep\sim e^{-\rho_\infty}$, its free energy should scale as $e^{\frac{d}{2}\rho_\infty}$, agreeing with (\ref{freenergy}). 

\section{Conclusions}
\label{sec:conclusion}
In this paper, we extensively studied holography with end-of-the-world (EOW) branes whose world volumes are flat, by extending the AdS/BCFT construction \cite{Takayanagi:2011zk,Fujita:2011fp}. These flat EOW branes have the peculiar feature that they intersect with the boundary of the global AdS on null surfaces. Thus, their gravity duals involve
conformal field theories with null boundaries which have not been studied well in the past. Moreover, the brane-world holography relates these to holography for flat spacetimes, which has  also been still far from complete understandings.

We classified interesting holographic setups with flat EOW branes into three classes: type I, II and III as in Fig.\ref{fig:setups}. The type I and II setup include a single EOW brane whose tension is negative $T=-\frac{d-1}{R}$ and positive $T=\frac{d-1}{R}$, respectively, in the $d+1$ dimensional bulk. The type III model is defined by the region between two flat EOW branes. The field theory dual of type I is given by a CFT confined inside a diamond with null boundaries. The boundaries look like making the final state projections. 
On the other hand, in the field theory dual of type II, the CFT on the diamond is coupled to a flat space gravity through its null boundaries. The degrees of freedom of the $d$ dimensional flat space gravity on the brane is dual to those on the null boundaries, whose part can be viewed as the Flat/Carrollian CFT (CCFT) correspondence.  In the type III model, the AdS boundary dual to the $d+1$ dimensional bulk is the $d-1$ dimensional null surfaces, which are the boundary of the diamond. Therefore, it singles out the part of the flat/CCFT holography from the type II construction, which can be thought of a version of the wedge holography \cite{Akal:2020wfl}.

By applying the AdS/BCFT description to these setups, we computed holographic entanglement entropy, correlation functions and on-shell actions. We found analytical expressions of holographic entanglement entropy both in type I and type II case, which are simply related to by the transformation $z_0\to -z_0$.
We noted that in the contributions from the disconnected geodesics in the type II case, there is a constant imaginary part in the entropy, which is similar to the time-like entanglement entropy \cite{Doi:2022iyj,Doi:2023zaf} and we gave a possible explanation in terms of a union of time-like and space-like geodesic. This implies that the dual density matrix is not hermitian. It would be intriguing to consider this from the viewpoint of the recently discussed connection between causal influence and non-hermitian density matrix \cite{Kawamoto:2025oko,Milekhin:2025ycm}.
Also this phenomenon may not be surprising as the dual of flat space gravity can be non-unitary, which was also pointed out in \cite{Pasterski:2022lsl,Ogawa:2022fhy}.
We leave more systematic studies of this interesting point for a future work. 
In both type I and II, there is a phase transition of holographic entanglement entropy between the contributions from the connected and disconnected geodesics by choosing the minimum of the real part of the entropy. We also computed the one-point functions and two-point functions in the type I and II case and found that they are related to each other via the transformation $z_0\to -z_0$. These provide strong predictions based on holography for CFTs with null boundaries. In the appendix A, we also performed field theory calculations by taking an infinite boost of a time-like boundary. This leads to a rather trivial result that the one-point function vanishes. It is possible that this corresponds to $z_0\to \infty$ limit of type I or type II setup, where the one-point functions also vanish. 

In the type III setup, by applying the wedge holography \cite{Akal:2020wfl}, we argued that the $d+1$ dimensional AdS bulk region between two EOW branes are dual to a $d-1$ dimensional Carrollian CFT (CCFT), which lives on the null surfaces, which are the edges of a diamond. Since the $d+1$ dimensional bulk gravity is dual to $d$ dimensional quantum gravity on the two EOW branes, this is essentially reduced to the Flat/CCFT correspondence. In other words, our AdS/BCFT approach gives a justification of the Flat/CCFT correspondence. For example, the analysis of holographic entanglement in this wedge holography reproduces the swing surface prescription of holographic entanglement in flat space holography \cite{Apolo:2020bld}. Moreover, the analysis of reduction of a scalar field in the $d+1$ dimensional bulk to a tower of massive scalar fields in the $d$ dimension brane shows that we can reduce the analysis of holographic correlation functions in the type III case to that in the Flat/CCFT correspondence. 

We also studied a Euclidean space version of flat space holography using the flat EOW brane by computing the on-shell action. This implies that the gravity dual of a flat space is given by a point-like theory because the flat EOW brane in the Euclidean AdS intersects with the AdS boundary only at a single point. It would be intriguing to compare this with the celestial holography \cite{deBoer:2003vf,Pasterski:2016qvg,Pasterski:2017kqt} 
and the Euclidean approach \cite{Li:2010dr} for the better understandings of our new holographic duality.

\section*{Acknowledgments}
This work is supported by MEXT KAKENHI Grant-in-Aid for Transformative Research Areas (A) through the ``Extreme Universe'' collaboration: Grant Number 21H05187. TT is also supported by Inamori Research Institute for Science, and by JSPS Grant-in-Aid for Scientific Research (B) No.~25K01000. PH is also supported by the NSFC special fund for theoretical physics No.\,12447108.
NO is supported by JSPS KAKENHI Grant Number JP24KJ1372. 
TW is supported by JSPS KAKENHI Grant Number JP25KJ1621. 


\appendix
\section{BCFT with a null boundary}
\label{app: BCFT}
In this appendix, we analysis the correlation functions in the two dimensional BCFTs with the boundary which is null.
\subsection{Free scalar model}
We firstly consider the free scalar model,
\begin{equation}
    S=\frac{1}{4\pi}\int d\sigma dt(-(\partial_t\phi)^2+(\partial_\sigma\phi)^2),
\end{equation}
on the cylinder $(t,\sigma)$ with length $L$ and also the boundary $B$. The variation gives the equation of motion 
\begin{equation}
    (-\partial_t^2+\partial_x^2)\phi=0,
\end{equation}
and also the boundary condition
\begin{equation}
    \text{Dirichlet:}\ \ \ \ \ \ \delta \phi|_B=0,
\end{equation}
or
\begin{equation}
   \text{Neumann:}\ \ \ \ \ \ \nabla_n\phi=0,
\end{equation}
where $n$ is a unit vector normal to $B$, and $\nabla_n$ is the derivative respect to that direction. The solution in terms of modes is
\begin{equation}
  \phi(t,\sigma)=\phi_0+\frac{4\pi}{L}\pi_0t+i\sum_{n\neq0}\frac{1}{n}a_ne^{-\frac{2\pi}{L}in(t-\sigma)}+i\sum_{m\neq0}\frac{1}{m}\bar a_me^{-\frac{2\pi}{L}im(t+\sigma)}.
\end{equation}
We can then impose the Dirichlet boundary condition at the boundary $B$
\begin{equation}
    \sigma=kt,
\end{equation}
which further imposes the constrains on the solution
\begin{equation}
    \bar a_m=-\frac{1-k}{1+k}a_n,\ \ m=\frac{1-k}{1+k}n.
\end{equation}
The green function is calculated from the mode sum
\begin{align}
    G_{cyl}(u_1,v_1;u_2,v_2)=&\log \left(1-e^{-\frac{2 i \pi  (k u_1+k
   v_2+u_1-v_2)}{(k+1) L}}\right)+\log \left(1-e^{\frac{2 i
   \pi  (k u_2+k v_1+u_2-v_1)}{(k+1) L}}\right)\nonumber\\
   &-\log
   \left(1-e^{\frac{2 i \pi  (k-1) (v_1-v_2)}{(k+1) L}}\right)-\log
   \left(1-e^{-\frac{2 i \pi  (u_1-u_2)}{L}}\right).
\end{align}
We can go to the plane limit $L\rightarrow\infty$, and calculate the one-point and two-point functions of the scalar operator $A=:\partial_u\phi\partial_v\phi
:$ with $(h,\bar h)=(1,1)$,
\begin{equation}
    \langle A(u,v)\rangle=\frac{1-k^2}{((k+1) u+(k-1) v)^2},
\end{equation}
\begin{equation}
    \langle A(u_1,v_1)A(u_2,v_2)\rangle=\frac{(k-1)^2}{\left(\frac{(k-1) v_2}{k+1}+u_1\right)^2 ((k+1)
   u_2+(k-1) v_1)^2}+\frac{1}{(u_1-u_2)^2
   (v_1-v_2)^2},
\end{equation}
where
\begin{equation}
    u=t-x,\ \ v=t+x.
\end{equation}
In the case where the boundary is null, we can take $k=\pm1$ and have 
\begin{equation}\label{eq:1 2 pt of A}
        \langle A\rangle=0,\ \  \langle AA\rangle=\frac{1}{(u_1-u_2)^2
   (v_1-v_2)^2}.
\end{equation}
This provides an example of the correlation functions in the BCFTs with null boundary, where the one-point function vanishes and the two-point function is the same as that in the CFT without the boundary. The discussion with the Neumann boundary condition is similar and the direct calculation shows that the one-point and two-point function of $A$ diverges as $\frac{1}{k^2-1}$ and $\frac{1}{(k^2-1)^4}$ as $k\to\pm1$. We can normalize $A$ so that it gives the same result as \eqref{eq:1 2 pt of A}.
\subsection{Analytical continuation and conformal map}
We can also consider the one-point function from the analytical continuation and conformal map. In the Euclidean signature, on the plane $(x,\tau)$ with the boundary $x=k\tau$, the scalar $O$ with $(h,h)$ has the one-point function
\begin{equation}
    \langle O\rangle=\frac{e^{2 ih\arctan k}}{\left(
   z e^{2 i \arctan k}-\bar z\right)^{2h}},
\end{equation}
where we used the transformation
\begin{equation}
    z\to e^{i\arctan k}z,\ \ \bar z\to e^{-i\arctan k}\bar z,
\end{equation}
from the upper half plane. For the null boundary, we should consider $k\to\pm i$ so that the boundary is $x=\mp t$ in the Lorentz signature and the one-point function reads
\begin{equation}
    \langle O\rangle_{k=\pm i}=0.
\end{equation}
This also gives the vanishing one-point function in the BCFT if the boundary is null.
With the two-point function of two identical scalar with scaling dimension $\Delta=2h$ on the upper half plane,
\begin{equation}
    \langle O_1(z_1,\bar z_1)O_2(z_2,\bar z_2)\rangle_{UHP}=\qty(\frac{\eta}{(z_1-\bar z_1)(z_2-\bar z_2)})^\Delta F(\eta),
\end{equation}
where $\eta=\frac{(z_1-\bar z_1)(z_2-\bar z_2)}{(z_1-\bar z_2)(z_2-\bar z_1)}$ is the cross ratio, we have the two-point function for arbitrary $k$,
\begin{equation}
    \langle O_1(z_1,\bar z_1)O_2(z_2,\bar z_2)\rangle_{k}=\qty(\frac{\eta'}{(e^{i\arctan k}z_1-e^{-i\arctan k}\bar z_1)(e^{i\arctan k}z_2-e^{-i\arctan k}\bar z_2)})^\Delta F(\eta'),
\end{equation}
where
\begin{equation}
    \eta'=\frac{(e^{i\arctan k}z_1-e^{-i\arctan k}\bar z_1)(e^{i\arctan k}z_2-e^{-i\arctan k}\bar z_2)}{(e^{i\arctan k}z_1-e^{-i\arctan k}\bar z_2)(e^{i\arctan k}z_2-e^{-i\arctan k}\bar z_1)}.
\end{equation}
For the null boundary with $k\to\pm i$, we find $\eta'\to1$ which indicates that the bulk channel is dominant and gives the same two-point function as that in the CFT without boundary,
\begin{equation}
   \lim_{k\to\pm i}\langle O_1(z_1,\bar z_1)O_2(z_2,\bar z_2)\rangle_{k}=\langle O_1(z_1,\bar z_1)O_2(z_2,\bar z_2)\rangle_{\text{CFT}}.
\end{equation}
\subsection{Moving mirror} 
\begin{figure}[ttt]\label{fig: mirror}
		\centering
		\includegraphics[width=6cm]{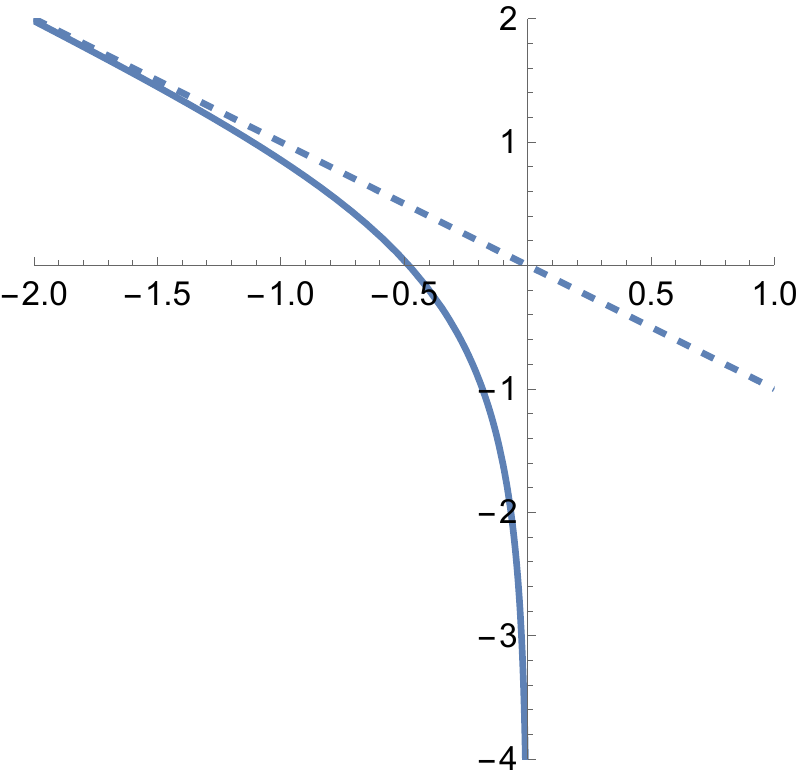}
		\caption{The solid line is the location of the moving mirror, while the dashed line is the null boundary.} 
\end{figure}

The null boundary effect can also be considered in the infinitely boosted moving mirror setup as shown in Fig.\eqref{fig: mirror} .  A static mirror is located at 
\begin{equation}\label{eq: static mirror}
    \tilde{x}=0,
\end{equation}
in the $(\tilde u=\tilde t-\tilde x,\tilde v=\tilde t+\tilde x)$ plane. A conformal transformation \cite{Akal:2022qei}
\begin{equation}
    \tilde u =p(u),\ \ \tilde v=v,
\end{equation}
maps the boundary \eqref{eq: static mirror} to 
\begin{equation}
    p(u)=v,
\end{equation}
in the $(u=t-x,v=t+x)$ plane. With a special choice of the map
\begin{equation}
    p(u)=-\log(1+e^{-u}),
\end{equation}
the profile of the boundary is described by
\begin{equation}
    t=\log(-2\sinh x)=\log(e^{-x}-e^x).
\end{equation}
As $t\to-\infty$, it becomes the usual time-like boundary
\begin{equation}
    x=0.
\end{equation}
As $t\to\infty$, it becomes null
\begin{equation}
    t=-x.
\end{equation}
Then, we can calculate the one-point function of a scalar $O$ with $(h,h)$ in the $(u,v)$ plane from the conformal map, and get
\begin{equation}
    \langle O(u,v)\rangle=\frac{p'(u)^h}{(p(u)-v)^{2h}}.
\end{equation}
For fixed $v$, we can go to the infinitely boosted region from the $u\to\infty$ limit and have
\begin{equation}
    \lim_{u\to\infty}\langle O(u,v)\rangle=0.
\end{equation}
Besides, in the $(t,x)$ coordinate,
\begin{equation}
    \langle O(t,x)\rangle=\left(\frac{1}{e^{t-x}+1}\right)^h \left(-\log
   \left(e^{x-t}+1\right)-t-x\right)^{-2 h},
\end{equation}
so that in the limit $t\to\infty$, we have
\begin{equation}
    \lim_{t\to\inf}\langle O(t,x)\rangle=0,
\end{equation}
with the null boundary and
\begin{equation}
    \lim_{t\to-\infty}\langle O(t,x)\rangle=\frac{1}{(2x)^{2h}},
\end{equation}
in the limit $t\to-\infty$ with the time-like boundary.
Then, the similar calculation gives the two-point function with the presence of the mirror,
\begin{equation}
    \langle O_1(u_1,v_1)O_2(u_2,v_2)\rangle_{\text{mirror}}=\qty(\frac{\eta_p\sqrt{p'(u1)p'(u2)}}{(p(u_1)-v_1)(p(u_2)-v_2)})^\Delta F(\eta_p),
\end{equation}
where the cross ratio is
\begin{equation}
    \eta_p=\frac{(p(u_1)-v_1)(p(u_2)-v_2)}{(p(u_1)-v_2)(p(u_2)-v_1)}.
\end{equation}
Again, we find that in both $t_{1,2}\to\infty$ and $u_{1,2}\to\infty$ limit, $\eta_p\to1$ so that the bulk channel is dominant.

\bibliographystyle{JHEP}
\bibliography{FlatEOW}


\end{document}